\pdfoutput=1 

\documentclass[a4paper,11pt]{article}
\usepackage{tabularx} 
\usepackage{amsmath}  
\usepackage{amssymb}  
\usepackage{graphicx} 
\usepackage[margin=1in,letterpaper]{geometry} 
\usepackage[final]{hyperref} 
\hypersetup{
	colorlinks=false,       
	linkcolor=blue,        
	citecolor=blue,        
	filecolor=magenta,     
	urlcolor=blue         
}

\usepackage[affil-it]{authblk}  

\usepackage[font=footnotesize,labelfont=bf]{caption} 

\usepackage{physics} 
\usepackage{slashed} 

\usepackage{xcolor} 
\usepackage{xfrac}
\usepackage{float} 
\usepackage{tabularray} 

\usepackage{subcaption}
\usepackage{wrapfig}

\usepackage{bbold}
\usepackage{bm}
\usepackage{dsfont}
\usepackage[normalem]{ulem}

\usepackage{titlesec}
\titleformat*{\subsection}{\bfseries\boldmath}

\newsavebox{\figbox}

\renewcommand{\O}{\mathcal{O}}

\newcommand{\PL}{\text{P}_\text{L}}
\newcommand{\PR}{\text{P}_\text{R}}

\renewcommand{\Tr}[1]{\text{Tr}\left\{#1\right\}}

\newcommand{\lambar}{\bar\lambda}

\newcommand{\MS}{$\overline{\text{MS}}$}

\usepackage[style=numeric-comp, backend=bibtex, maxnames=4, sorting=none]{biblatex} 
\title{\bfseries\boldmath The dipole ellipse from vector-like leptons at next-to-leading order }

\author[1]{Kilian Möhling \thanks{kilian.moehling@tu-dresden.de}}
\affil[1]{Institut für Kern- und Teilchenphysik, TU Dresden, Zellescher Weg 19, 01069 Dresden, Germany}
\date{}

\begin{document}
\maketitle


\begin{abstract}
	We consider the SM extended by a 4th generation of vector-like leptons. This model generates strongly correlated
	contributions to the lepton masses, Higgs couplings and dipole moments. At leading order 
	the Higgs coupling restricts the magnetic and electric dipole moments to an ellipse. 
	Here we compute the one-loop corrections to the Higgs coupling and
	two-loop corrections to the dipole moments to obtain the correlation at next-to-leading order using the framework of SMEFT.
	In addition we resum the leading logarithmic corrections from SM couplings through the SMEFT renormalisation group equations.
	With these results we compute the RG improved correlation at NLO and
	discuss the implications for the muon, electron and tau correlations. Specifically, for the muon we find that the corrections
	contract the ellipse such that a larger part of parameter space becomes compatible with the updated SM value of the muon magnetic moment.
\end{abstract}

\setcounter{tocdepth}{2}
\tableofcontents

\section{Introduction}
\label{sec:intro}

Vector-like leptons (VLL) naturally arise in a variety of UV completions \cite{delAguila:1982fs}, grand-unified theories \cite{Gursey:1975ki,Langacker:1988ur,Hewett:1988xc,Dermisek:2012ke,Dermisek:2012as,Dorsner:2014wva,Bhattacherjee:2017cxh,Kowalska:2019qxm} or models with extra dimensions
\cite{Huber:2000ie,Gherghetta:2000qt,Agashe:2008fe}. 
In the past, the phenomenology of VLL has been surveyed extensively~\cite{Frampton:1999xi,Ellis:2014dza,Ishiwata:2015cga,Raby:2017igl,Arkani-Hamed:2021xlp,Baspehlivan:2022qet}. 
Through Yukawa-type mixing with the SM leptons, VLL affect a broad range of observables including Higgs decays and collider signatures~\cite{Ishiwata:2011hr,Kearney:2012zi,Falkowski:2013jya,Daberstiel:2026wxe}, 
lepton flavour violation~\cite{Goudelis:2011un,Ishiwata:2013gma,Poh:2017tfo,Daberstiel:2026wxe}, 
electroweak precision observables~\cite{Crivellin:2020ebi,delAguila:2008pw}, 
lepton electric dipole moments (EDMs)~\cite{Hamaguchi:2022byw,Dermisek:2022aec,Dermisek:2023nhe} 
and lepton anomalous magnetic moments~\cite{Kannike:2011ng,Dermisek:2013gta,Freitas:2014pua,Athron:2021iuf,Athron:2025ets}.

Here we focus specifically on a fourth generation of VLL, i.e. the Standard Model (SM) extended by one Dirac SU$(2)_W$ singlet $E$ and one doublet $L$ with
the same hypercharges as the corresponding chiral SM fields.
A particularly distinctive feature arises when the SM leptons mix with both representations of the VLL. 
In this case the VLL Yukawa couplings provide a new source of lepton chiral symmetry breaking, independent of the SM Yukawa coupling.
This, in turn, results in strongly correlated contributions to the lepton mass, Higgs coupling and dipole moments \cite{Kannike:2011ng,Kearney:2012zi,Dermisek:2013gta,Falkowski:2013jya,Crivellin:2021rbq,Mohling:2024qvk}. 
Bounds on the lepton--Higgs coupling then lead to an elliptical constraint on the dipole moments \cite{Dermisek:2022aec,Dermisek:2023nhe} 
that can be probed with current and upcoming experiments.

For the muon, the Fermilab $g-2$ experiment has recently published its final result \cite{Muong-2:2025xyk}. At the same time, 
the SM prediction has been updated by the theory initiative \cite{Aliberti:2025beg} (WP25) 
and is now in good agreement with the Fermilab measurement, which motivates
revisiting some of the previous analyses to reevaluate the impact of the new result \cite{Athron:2025ets}.
In the near future substantial improvements in the experimental sensitivity of the muon EDM are also expected \cite{Sakurai:2022tbk}
that will provide the missing ingredient to probe the correlations predicted by the VLL. 
On the high-energy front, evidence for $h\to\mu^+\mu^-$ has been reported by CMS and ATLAS~\cite{CMS:2020xwi,CMS:2022dwd,ATLAS:2025coj}, 
giving a first bound on the muon--Higgs coupling $\lambda_{\mu\mu}$. 

For the electron, the magnetic moment is measured to $0.13$\,ppt~\cite{Fan:2022eto}, and the electron EDM is bounded at
$|d_e|<4.1\times10^{-30}\,e\,$cm~\cite{Roussy:2022cmp}. However, the electron-Higgs coupling $\lambda_{ee}$ 
is currently very weakly constrained by direct searches for $h\to e^+e^-$~\cite{ATLAS:2019old,CMS:2022urr}, 
leaving a lot of room for new physics~\cite{Erdelyi:2025axy}. 
For the $\tau$ the situation is reversed. The Higgs coupling is measured at the ten-percent level~\cite{ATLAS:2022vkf,CMS:2022dwd}, 
while the dipole moments are only loosely bounded, e.g.\ from $\gamma\gamma\to\tau\tau$ production at the LHC~\cite{ATLAS:2022ryk,CMS:2024qjo}. 
With substantial improvements expected on the high-energy side at the HL-LHC and future colliders~\cite{Cepeda:2019klc,deBlas:2019rxi,dEnterria:2021xij,FCC:2025lpp}, 
the dipole--Higgs correlation will be tested in all three generations. 
It is therefore important that the theory prediction for the correlation be reliable beyond leading order.

This question was partially addressed in Ref.~\cite{Mohling:2024qvk} where an on-shell renormalisation scheme for the 
VLL model was set up and the one-loop corrections to the muon--Higgs coupling were computed in the full theory. 
The leading-order correlation (in case of CP-conservation and $\lambda_{\mu\mu}\approx -\lambda_{\mu\mu}^\text{SM}$) 
implies a fixed value $\Delta a_\mu\approx 23\times10^{-10}$, in agreement with the deviation based on the previous SM value from the 
2020 white paper (WP20) \cite{Aoyama:2020ynm}. 
The next-to-leading order (NLO) corrections to the Higgs coupling were found to reduce this value significantly, to the range $(10\ldots18)\times10^{-10}$
overlapping with the updated WP25 value at 2$\sigma$. 
This suggests that higher-order effects can substantially modify the correlation and, in particular, the conclusions drawn from the updated value of $\Delta a_\mu$.

However, in Ref.~\cite{Mohling:2024qvk} the dipole moments were treated at leading order, so the NLO correlation was incomplete. 
Since the leading dipole contribution is itself a one-loop effect, the missing piece are the two-loop corrections. 
In this work we compute this correction and complete the NLO prediction for the correlation.

For VLL masses well above the electroweak scale, the natural framework for such a calculation is SMEFT~\cite{Buchmuller:1985jz,Grzadkowski:2010es}. 
At tree-level the VLL generate the dimension-six operator $\mathcal{Q}_{e\Phi}$, that modifies the relation between lepton mass and Higgs coupling,
as well as $\mathcal{Q}_{eB}$ and $\mathcal{Q}_{eW}$ at one-loop both of which contribute to the lepton dipole moments~\cite{Kannike:2011ng,deBlas:2017xtg,Chala:2020odv,Crivellin:2020ebi,Dermisek:2023nhe,Erdelyi:2025axy}. 
 
The EFT separates the calculation into a matching condition at the VLL mass scale and a renormalisation group (RG) evolution down to the electroweak and lepton mass scales. This approach has two practical advantages over a full calculation in the mass basis:
(i) the actual two-loop (matching) calculation is significantly simpler due to the fact that only vacuum integrals with (at most) two mass scales appear and
(ii) large logarithmic corrections from SM couplings are resummed rather than computed at fixed order leading to an improved accuracy of the result.
The complete one-loop RGEs are known for both SMEFT~\cite{Jenkins:2013zja,Jenkins:2013wua,Alonso:2013hga} and LEFT~\cite{Czarnecki:1995sz,Degrassi:1998es,Jenkins:2017dyc}, 
where the former are used to resum logarithms of the form $\ln(M_\text{VLL}/M_Z)$ while the latter reproduce the large universal QED logarithm in the dipole moments from running below
the EW scale~\cite{Czarnecki:1995sz,Degrassi:1998es}. 

At NLO, $C_{e\Phi}$ requires matching at one-loop and the dipole coefficients at two-loop. 
One-loop matching is by now largely automated~\cite{Fuentes-Martin:2022jrf,Carmona:2021xtq,Guedes:2023azv}. 
However, while functional methods have recently been extended to two loops~\cite{Fuentes-Martin:2023ljp}, 
computing two-loop matching results for specific models remains involved (see e.g. Ref.~\cite{Panico:2018hal} for a related two-loop EFT analysis of EDMs) and automated tools are not yet publicly available.

In this work we compute the NLO matching conditions for $C_{e\Phi}$, $C_{eB}$ and $C_{eW}$ at $\mathcal{O}(\lambda^5_\text{VLL})$ in the VLL Yukawa couplings, for arbitrary VLL masses $M_L$ and $M_E$. 
In addition to the genuine two-loop matching diagrams two more contributions have to be taken into account. 
The first is the canonical normalisation of the EFT fields. 
The second are the threshold corrections to the Higgs mass parameter, which naively enter at LO with an additional mass suppression (dimension 8).
However, since they are loop suppressed but proportional to the heavy VLL masses, they instead enter at NLO and dimension 6, i.e. the same order
as the genuine two-loop diagrams.

After combining the matching results with the resummation of the leading logarithms, we study how the corrections distort the elliptic correlation 
between the dipole moments and the Higgs coupling for the muon, electron and $\tau$.

The paper is organized as follows. In Sec.~\ref{sec:LO-correlation} we define the model and review the leading-order correlation. Sec.~\ref{sec:NLO} presents the NLO matching conditions, including the field normalisation, the Higgs mass parameter contributions and genuine two-loop diagrams. Sec.~\ref{sec:resummed-obs} discusses the calculation of the RG-improved observables. 
We present the numerical results in Sec.~\ref{sec:numeric} and conclude in Sec.~\ref{sec:conclusions}. The complete mass dependence of the matching coefficients and the analytic expressions for the two-loop master integrals are collected in Apps.~\ref{app:complete-result} and~\ref{app:loop-functions}.

\section{Model definition and leading order correlation}\label{sec:LO-correlation}
We consider the Standard Model (SM) extended by a 4th generation of vector-like leptons in the same
representation as the SM leptons. The relevant Lagrangian for the interactions between the VLL and SM leptons 
is given by
\begin{align}
	-\mathcal{L} \supset y_i \,\big(\overline{l_{Li}}\Phi\big) e_{Ri} + \lambda^L_i \big(\overline{L}\Phi\big) e_{Ri} + \lambda^E_i \big(\overline{l_{Li}}\Phi\big) E 
	+ \big(\overline{L}\Phi\big)\Big[\lambda\PR + \lambar^* \PL \Big] E + h.c.
\end{align}
where $L$ and $E$ denote the VLL doublet and singlet with hypercharges $-\frac12$ and $-1$ respectively, while 
$l_{Li}$ and $e_{Ri}$ denote the left- and right-handed SM lepton fields and $\Phi$ the SM Higgs doublet with $\ev{\Phi}=(0,v)^T$
and $v\approx 174$ GeV. 
In the following we collectively refer to the new VLL couplings by $\lambda_\text{VLL}$ and the SM couplings as $g_\text{SM}$.

Both VLL carry a gauge-invariant Dirac mass respectively denoted by $M_L$ and $M_E$. These masses are entirely independent of the usual Higgs mechanism for the lepton mass generation
and in principle can be arbitrarily heavy. For masses significantly above the EW scale the VLL can be integrated out in the unbroken phase,
yielding matching conditions for the SMEFT Wilson coefficients.

\begin{figure}
	\centering
	\begin{subfigure}{.3\textwidth}
		\includegraphics[width=.8\textwidth]{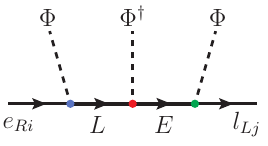}
	\end{subfigure}\hspace{2cm}
	\begin{subfigure}{.3\textwidth}
		\includegraphics[width=.8\textwidth]{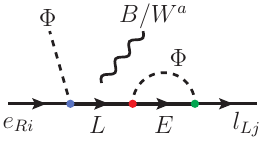}
	\end{subfigure}
	\caption{Diagrams contributing to the leading order matching conditions for $C_{e\Phi}$ (left) and $C_{eB}$ or $C_{eW}$ (right).}
	\label{fig:SMEFT-LO-diagrams}
\end{figure}

Here we are in particular interested in the SMEFT operators that generate the leading corrections to the lepton Higgs coupling and dipole moments.
They are given by
\begin{align}
	-\mathcal{L}_\text{SMEFT} \supset C_{\underset{ji}{e\Phi}} \big(\overline{l_{Lj}}\Phi\big) e_{Ri} |\Phi|^2
	+ C_{\underset{ji}{eB}} \big(\overline{l_{Lj}}\Phi\big)\sigma^{\mu\nu} e_{Ri} B_{\mu\nu}
	+ C_{\underset{ji}{eW}} \big(\overline{l_{Lj}}\sigma_a\Phi\big)\sigma^{\mu\nu} e_{Ri} W^a_{\mu\nu}
	+ h.c.
\end{align}
where $B_{\mu\nu}$ and $W^a_{\mu\nu}$ denote the field strength tensors of the EW gauge bosons, $\sigma_a$ the Pauli matrices
and $\sigma^{\mu\nu} = \frac{i}{2}[\gamma^\mu,\gamma^\nu]$.
This class of operators is distinct in that each of them provides a separate breaking of the SM lepton chiral symmetry
in addition to the renormalisable SM Yukawa coupling $y_i$.
As a consequence the contributions to these Wilson coefficients are proportional to the following two independent spurions breaking the SM lepton chiral symmetry
\begin{align}
	\frac{\lambda^L_i \lambar \lambda^E_j}{M_E M_L}, \qquad \text{and} \qquad \lambda^L_i \lambda^* \lambda^E_j.
\end{align}
At leading order, only the first combination actually appears, resulting in a strong correlation between the Wilson coefficients and observables predicted by them \cite{Kearney:2012zi,Dermisek:2013gta,Falkowski:2013jya,Dermisek:2022aec,Dermisek:2023nhe,Mohling:2024qvk}. 
This correlation can also be seen diagrammatically in Fig.~\ref{fig:SMEFT-LO-diagrams}. The resulting leading order matching conditions read
\begin{align}\label{eq:LO-matching}
	C_{\underset{ji}{e\Phi}} = \frac{\lambda^L_i\lambda^E_j\lambar}{M_E M_L}, \qquad
	C_{\underset{ji}{eB}} = \frac{3 g_Y }{128\pi^2} \frac{\lambda^L_i\lambda^E_j \lambar}{M_E M_L} ,\qquad
	C_{\underset{ji}{eW}} = \frac{ g_W }{128\pi^2} \frac{\lambda^L_i\lambda^E_j \lambar}{M_E M_L}.
\end{align}
While these leading order conditions only depend on the renormalisation scale $\bar\mu$ implicitly through the VLL Yukawa couplings
and Dirac masses, here we already assume the couplings are evaluated at the matching scale $\bar\mu_\Lambda=M_L$, in order to avoid large
logarithmic corrections in the NLO matching condition. This also allows us to resum the leading logarithmic SM corrections to these matching
conditions which is discussed in more detail in Sec.~\ref{sec:resummed-obs}. After EWSB, the coefficient $C_{e\Phi}$ generates
a correction to the physical lepton mass $m_{\ell_i}$ and Higgs coupling $\lambda_{\ell_i}$ which, crucially, differ by a combinatorial factor
such that $C_{e\Phi}$ breaks the usual relation $\lambda_{\ell_i}=m_{\ell_i}/v$ and instead gives
\begin{subequations}\label{eq:lepton-Higgs-LO}
	\begin{align}
	m_{\ell_i} &= y_{\ell_i} v + C_{\underset{ii}{e\Phi}} v^3, \\
	\lambda_{\ell_i} &= y_{\ell_i} + 3 C_{\underset{ii}{e\Phi}} v^2 = \frac{m_{\ell_i}}{v} + 2 C_{\underset{ii}{e\Phi}} v^2.
	\end{align}
\end{subequations}
Here in principle an additional relative phase between the SMEFT Yukawa coupling $y_{\ell_i}$ and $C_{e\Phi}$ could appear \cite{Dermisek:2023nhe},
but (without loss of generality) can be absorbed into a redefinition of the UV Yukawa coupling $y_i$\footnote{Note that the SMEFT Yukawa coupling 
	$y_{\ell_i}$ differs from the UV coupling $y_i$ by one-loop threshold corrections which, however, drop out in the physical correlations.
	In a calculation in the full model this fact is slightly obscured but can be seen e.g. using on-shell renormalisation \cite{Mohling:2024qvk},
	while in the SMEFT setup it is immediately apparent after eliminating $y_{\ell_i}$ in favour of the physical lepton masses.}.
The lepton dipole moments are given at leading order by
\begin{align}\label{eq:lepton-dipole-LO}
	\Delta a_{\ell_i} &= -\frac{4 m_{\ell_i} v}{e}\Re{C_{\underset{ii}{e\gamma}}}, \qquad
	d_{\ell_i} = 2v\,\Im{C_{\underset{ii}{e\gamma}}}
\end{align}
where $C_{e\gamma} \equiv c_W C_{eB} - s_W C_{eW}$ and $s_W$ ($c_W$) denotes the (co)sine of the weak mixing angle. 
According to the leading order matching conditions Eq.~\eqref{eq:LO-matching}
\begin{align}\label{eq:def-kappa}
	C_{\underset{ji}{e\gamma}} = \frac{e}{\kappa} C_{\underset{ji}{e\Phi}} \qquad \text{with} \qquad 
	\kappa \approx 64\pi^2
\end{align}
which implies a direct correlation between the Higgs signal strength $\mu(h\to\ell_i\overline{\ell_i}) \approx |\lambda_{\ell_i}/\lambda_{\ell_i}^\text{SM}|^2 \equiv R_{\ell_i\ell_i}$
and the lepton dipole moments given by
\begin{align}
	R_{\ell_i\ell_i}\Big|_\text{LO} = \left(1 - \frac{\Delta a_{\ell_i}}{2\omega_i}\right)^2 + \left(\frac{m_{\ell_i} d_{\ell_i}}{e\omega_i}\right)^2,
\end{align}
where $\omega_i = m_{\ell_i}^2/(64\pi^2 v^2)$. Thus, the experimental constraint on $R_{\ell_i\ell_i}$ restricts the model to an ellipse 
in the $\Delta a_{\ell_i}-d_{\ell_i}$ plane \cite{Dermisek:2022aec}. The NLO matching conditions for the Wilson coefficients modify the coefficient $\kappa$
\begin{align}\label{eq:kappa-NLO}
	\kappa = 64\pi^2 \Big(1 + \O(1\ell)\Big)
\end{align}
and in particular introduce a mass and coupling dependence together with a potential imaginary part $\Im\kappa\neq 0$. The effect is a shift
and rescaling of the elliptical constraint on the dipole moments.
\begin{align}
	R_{\ell_i\ell_i} = \left(\frac{\Re\kappa}{|\kappa|} - \frac{|\kappa|}{64\pi^2}\frac{\Delta a_{\ell_i}}{2\omega_i}\right)^2 + 
	\left(\frac{\Im\kappa}{|\kappa|} - \frac{|\kappa|}{64\pi^2} \frac{m_{\ell_i} d_{\ell_i}}{e\omega_i}\right)^2.
\end{align}
The relevant NLO correction to $\kappa$ comes from both the one-loop contribution to $C_{e\Phi}$ as well as the two-loop contributions
to $C_{e\gamma}$. In addition one also potentially has to take into account NLO SMEFT corrections to the observables which will be discussed
in Sec.~\ref{sec:resummed-obs}.

\section{Matching conditions at next-to-leading order}\label{sec:NLO}
In this section we compute the VLL NLO contributions to the matching conditions, i.e.
one-loop corrections to $C_{e\Phi}$ and two-loop corrections to $C_{e\gamma}$. Specifically,
we focus on the pure VLL corrections, i.e. the terms of $\O(\lambda_\text{VLL}^5)$. While we neglect terms of $\O(\lambda_\text{VLL}^3 g_\text{SM}^2)$ 
in the matching condition, the more important logarithmically enhanced
corrections of this form are taken into account via the SMEFT running when computing the observables
(see Sec.~\ref{sec:resummed-obs}).
With this approximation only diagrams with internal couplings of the Higgs doublet to at least one VLL contribute,
while diagrams with internal gauge bosons or SM Yukawa couplings are neglected.

The NLO result consists of both the direct contribution from the NLO amplitudes as well as indirect effects
from EFT field normalisation and the Higgs mass threshold corrections, all of which contribute at $\O(\lambda_\text{VLL}^5)$.
In the following we describe each calculation and collect the relevant results. \\

The amplitudes were generated using \texttt{FeynArts} \cite{Hahn:2000kx} and the corresponding model file using 
\texttt{FeynRules} \cite{Christensen:2008py,Alloul:2013bka}. The amplitude expansion and tensor reduction was performed using \texttt{FeynCalc}
\cite{Mertig:1990an,Shtabovenko:2016sxi,Shtabovenko:2020gxv,Shtabovenko:2023idz} and the IBP reduction with \texttt{FIRE} \cite{Smirnov:2019qkx,Smirnov:2025prc} via the \texttt{FeynHelpers} \cite{Shtabovenko:2016whf,Shtabovenko:2025lxq} interface.
Furthermore, the one-loop results were checked using \texttt{Matchete} \cite{Fuentes-Martin:2022jrf}.

\subsection{EFT Field normalisation}\label{sec:field-norm}
	The kinetic matching generically yields non-canonically normalized kinetic terms in the EFT and the required (finite) field redefinition
	in turn changes the values of all Wilson coefficients at NLO. Writing the EFT kinetic Lagrangian in the Green basis as
	\begin{align}
		\mathcal{L}_\text{EFT} \supset i (\mathds{1}+\delta Z_e)_{ji} \; \overline{e_{Rj}} \slashed{D} e_{Ri} 
		+ i(\mathds{1}+\delta Z_l)_{ji} \; \overline{l_{Lj}} \slashed{D} l_{Li}
		+ (1+\delta Z_\Phi) |D_\mu\Phi|^2
	\end{align}
	where the Hermitian lepton field normalisation constants (after \MS\ renormalisation) are given by\footnote{We ignore corrections
		from gauge couplings, therefore the EFT gauge bosons do not receive WF corrections.}
	\begin{align}
		\delta Z_{e,ji} &= \frac{ \lambda^L_i\lambda^{L*}_j}{32\pi^2} \Big[3 - 4\,L(M_L)\Big] \qquad \text{and}\qquad
		\delta Z_{l,ji}  = \frac{\lambda^E_i\lambda^{E*}_j}{64\pi^2} \Big[3 - 4\,L(M_E)\Big],
	\end{align}
	with $L(m^n) \equiv n\,\ln(m/\bar\mu)$. The Higgs doublet normalisation is given by (with $r=M_E^2/M_L^2$)
	\begin{align}
		\begin{split}
			\delta Z_\Phi &= \frac{|\lambda^E_k|^2}{32\pi^2} \Big[1-4\,L(M_E)\Big] 
			+ \frac{|\lambda^L_k|^2}{32\pi^2} \Big[1-4\,L(M_L)\Big]
			+ \frac{\Re{\lambda\lambar}}{8\pi^2} \frac{[1-r^2 + 2 r \ln(r)]\sqrt{r}}{(r-1)^3} \\
			& + \frac{|\lambda|^2+|\lambar|^2}{32\pi^2} \Bigg\{ 1-\frac{4 r}{(r-1)^2} - \frac{2r^2(r-3)}{(r-1)^3}\ln (r) - 4 \, L(M_L)\Bigg\},
		\end{split}
	\end{align}
	where the sum over $k$ is implied. Note that the apparent pole at $r=1$ cancels yielding a finite result in that limit.
	The shift in the Wilson coefficients resulting from the canonical normalisation is given by 
	\begin{subequations}\label{eq:WC-WF-shift}
		\begin{align}
			C_{\underset{ji}{e\Phi}} &\longrightarrow C_{\underset{ji}{e\Phi}} \Big[1 - \tfrac{3}{2}\delta Z_\Phi\Big] - \tfrac{1}{2} \delta Z_{l,jk} C_{\underset{ki}{e\Phi}} - \tfrac{1}{2} C_{\underset{jk}{e\Phi}} \delta Z_{e,ki} \\
			C_{\underset{ji}{eB}} &\longrightarrow C_{\underset{ji}{eB}} \Big[1 - \tfrac{1}{2}\delta Z_\Phi\Big] - \tfrac{1}{2} \delta Z_{l,jk} C_{\underset{ki}{eB}} - \tfrac{1}{2} C_{\underset{jk}{eB}} \delta Z_{e,ki}  \\
			C_{\underset{ji}{eW}} &\longrightarrow C_{\underset{ji}{eW}} \Big[1 - \tfrac{1}{2}\delta Z_\Phi\Big] - \tfrac{1}{2} \delta Z_{l,jk} C_{\underset{ki}{eW}} - \tfrac{1}{2} C_{\underset{jk}{eW}} \delta Z_{e,ki}
		\end{align}
	\end{subequations}
	Note that field redefinitions at dimension six, e.g. from the $D^4\Phi^2$ or $D^3\Psi^2$ Green basis operators in principle contribute 
	to the NLO matching of $C_{e\Phi}$ through the dim-4 tree-level coupling $y_{\ell_i}$. However, these contributions are additionally
	suppressed by the small SM Yukawa coupling compared to the $\O(\lambda_\text{VLL}^5)$ terms and are thus neglected in our approximation. 
	Similarly, the normalisation from the $D^2\Phi^4$ operators	can also be dropped for our purposes.

\subsection{Higgs mass parameter matching}\label{sec:Higgs-mass}
	The Higgs mass parameter receives corrections proportional to the square of the heavy VLL masses. These corrections
	appear in the matching calculations from the difference of the Higgs mass parameter of SMEFT and the UV theory
	\begin{align}
		\mu_\text{SMEFT}^2 = \mu_\text{UV}^2 + \Delta \mu^2.
	\end{align}
	where after \MS\ renormalisation
	\begin{align}
		\begin{split}
			\Delta\mu^2 &= \frac{M_L^2}{8\pi^2} \Bigg[ |\lambda^E_k|^2 r\Big(L(M_E^2)-1\Big) + 
			|\lambda^L_k|^2\Big(L(M_L^2) - 1\Big) \\
			&\hspace{-.5cm}+ 2\Re{\lambda\lambar}\sqrt{r} \, \frac{rL(M_E^2) - L(M_L^2)+1-r}{r-1}
			+ \Big(|\lambda|^2+|\lambar|^2\Big) \frac{r^2L(M_E^2) - L(M_L^2) + 1-r^2}{r-1}
			\Bigg].
		\end{split}
	\end{align}
	By definition the SMEFT parameter reproduces the correct Higgs mass $\mu_\text{SMEFT}^2\approx \frac{1}{2}M_h^2$. At the same time, $\Delta\mu^2$
	is proportional to $M_L^2$ and, despite the loop suppression, can easily exceed the EW scale. Consequently, $\mu^2_\text{UV}$ has
	to be adjusted to cancel these large corrections leading to the well known fine-tuning problem. 
	Specifically, $\mu^2_\text{UV} \simeq - \Delta\mu^2$ and therefore the formally dim-8 LO matching condition
	$\propto \mu^2_\text{UV}/M_L^2$ results instead in a dim-6 NLO contribution	
	\begin{subequations}\label{eq:Ceg-mu2}
		\begin{align}
			\Delta_{\mu^2} C_{\underset{ji}{eB}} &= \frac{g_Y}{32\pi^2} \frac{ \lambda^L_i \lambda^E_j \lambar}{M_E M_L}
			\frac{\mu^2_\text{UV}}{M_E^2} \bigg\{ 2r\, \Big[2L(M_L^2)-5\Big] - \Big[2\,L(M_E^2) - 5\Big]\bigg\} 
			\\[.2cm]
			\Delta_{\mu^2} C_{\underset{ji}{eW}} &= \frac{g_W}{16\pi^2} \frac{ \lambda^L_i \lambda^E_j \lambar}{M_E M_L}
			\frac{\mu^2_\text{UV}}{M_E^2} \Big[5 - 2\,L(M_E^2)\Big].
		\end{align}
	\end{subequations}
	The corresponding diagrams are the same as in Fig.~\ref{fig:Ceg-2loop-CT} (right) where the insertion is identified with $\mu^2_\text{UV}$.
	Because of the massless Higgs lines, these diagrams are also IR divergent which however 
	cancels with the remaining IR divergence of the two-loop amplitudes after sub-renormalisation.

\subsection{$C_{e\Phi}$ at one-loop order}

	Next we look at the NLO (one-loop) matching of $C_{e\Phi}$.	The one-loop corrections can be written as
	\begin{align}\label{eq:CeH-1loop}
		\Delta_{1\ell}C_{\underset{ji}{e\Phi}} = - \frac{\lambda^L_i\lambda^E_j\lambar}{16\pi^2 M_E M_L}  \bigg\{ c_1 |\lambda^E_k|^2 + c_2 |\lambda^L_k|^2 
		+ c_3 \Big(|\lambda|^2 + |\lambar|^2\Big) + c_4 \lambda^* \lambar^* + c_5 \lambda\lambar
		\bigg\}
	\end{align}
	where (including the wave-function normalisation contribution Eq.~\eqref{eq:WC-WF-shift}) we find in the limit $M_E=M_L$ and at the matching scale $\bar\mu_\Lambda=M_L$
	\begin{align}
		c_1 = - \tfrac{17}{8}, \qquad 
		c_2 = - \tfrac{5}{2}, \qquad
		c_3 = c_4 = -\tfrac{1}{2}, \qquad
		c_5 = \tfrac{15}{2}
	\end{align}
	The complete result for $M_E\neq M_L$ is listed in App.~\ref{app:complete-result}.
	To compare with \texttt{Matchete}, similar to the Higgs mass parameter the UV Yukawa couplings first have to be systematically eliminated in favour of the $y_{\ell_i}$. 
	Since the Yukawa term contributes to $\mathcal{Q}_{e\Phi}$ at tree-level 
	via $\mathcal{Q}_{\Phi e}$ and $\mathcal{Q}_{\Phi l}^{(1,3)}$ \cite{Grzadkowski:2010es}, after the EOM reduction the above result agrees with Matchete only after inserting the explicit 
	one-loop threshold corrections from the dimension-4 matching. \\

\subsection{$C_{e\gamma}$ at two-loop order}
	\begin{figure}
		\centering
		\includegraphics[width=\textwidth]{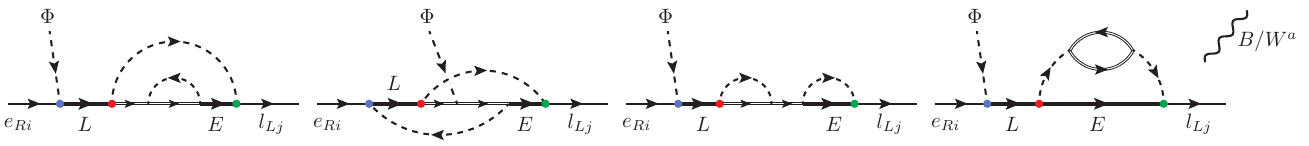}
		\caption{Example two-loop diagrams contributing to the dipole matching. 
			In total there are 120 diagrams of this form contributing to $C_{eB}$ and 80 contributing to $C_{eW}$.
			The hollow lines correspond to either VLL or SM leptons.}
		\label{fig:Ceg-2loop}
	\end{figure}
	\begin{figure}
		\centering
		\includegraphics[width=\textwidth]{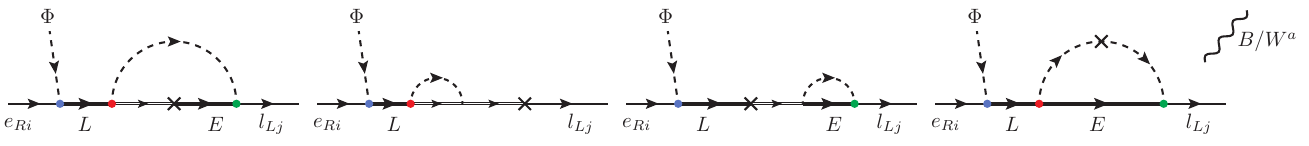}
		\caption{Example two-loop counter-term diagrams contributing to the dipole matching. 
			In total there are 110 counterterm diagrams contributing to $C_{eB}$ and 68 contributing to $C_{eW}$.
			The rightmost class of diagram (with the boson attached to an internal line)
			can also be interpreted as the formally dim-8 $\mu^2$ contribution explained in the text.}
		\label{fig:Ceg-2loop-CT}
	\end{figure}
	Finally, we discuss the genuine two-loop matching calculation of the dipole operators.	
	The relevant two-loop diagram classes are illustrated in Fig.~\ref{fig:Ceg-2loop} and 
	the corresponding counterterm diagrams sub-renormalising the amplitudes in Fig.~\ref{fig:Ceg-2loop-CT}.
	After sub-renormalisation the two-loop diagrams are UV finite, but still contain an IR divergence that cancels against
	the $\mu^2$ contribution discussed above. 
	In contrast to the NLO $C_{e\Phi}$, here the two-loop diagrams also generate corrections proportional to the second chirality breaking 
	order parameter $\lambda^E_i\lambda^L_j \lambda^*$, independent of $\lambar$. For simplicity, here we only list the result for $C_{e\gamma}=c_W C_{eB} - s_W C_{eW}$ 
	in the limit $M_E=M_L$, which can be written as
	\begin{align}\label{eq:Ceg-2loop}
		\begin{split}
			\Delta_{2\ell} C_{\underset{ji}{e\gamma}} = \frac{e\lambda^L_i \lambda^E_j}{3(8\pi)^4 M_E M_L} \bigg\{
			& \lambar \,\; \Big( a_1 |\lambda^E_k|^2 + a_2 |\lambda^L_k|^2 + a_3 |\lambda|^2 + a_4 |\lambar|^2 + a_5 \lambar\lambda \Big) \\
			+&\lambda^* \Big( \tilde{a}_1 |\lambda^E_k|^2 + \tilde{a}_2 |\lambda^L_k|^2 + \tilde{a}_3 |\lambda|^2 + \tilde{a}_4 |\lambar|^2 + \tilde{a}_5 \lambar^*\lambda^* \Big)
			\bigg\}
		\end{split}
	\end{align}
	where the coefficients $a_i$ and $\tilde{a}_i$ at the matching scale $\bar\mu_\Lambda=M_L$ are given by
	\begin{subequations}
		\begin{alignat}{3}
			a_1 &= -\tfrac{40 G}{\sqrt{3}}+6 \pi ^2-\tfrac{671}{4},      &\qquad&\qquad& \tilde{a}_1 &= \tfrac{2 \pi ^2}{3}-\tfrac{8 G}{\sqrt{3}},  \\
			a_2 &= -\tfrac{40 G}{\sqrt{3}}+6 \pi ^2-188,                &\qquad&\qquad& \tilde{a}_2 &= \tfrac{2 \pi ^2}{3}-\tfrac{8 G}{\sqrt{3}},  \\
			a_3 &= \tfrac{848 G}{\sqrt{3}}-\tfrac{967 \pi ^2}{54}-525,	  &\qquad&\qquad& \tilde{a}_3 &= -\tfrac{16 G}{\sqrt{3}}+\tfrac{289 \pi ^2}{108}-16, \\
			a_4 &= \tfrac{792 G}{\sqrt{3}} -\tfrac{1721 \pi ^2}{108}-513,         &\qquad&\qquad& \tilde{a}_4 &= \tfrac{848 G}{\sqrt{3}}-\tfrac{211 \pi ^2}{54}-543, \\
			a_5 &= \tfrac{848 G}{\sqrt{3}}-\tfrac{1721 \pi ^2}{108}-303, &\qquad&\qquad& \tilde{a}_5 &= -\tfrac{8 G}{\sqrt{3}}+\tfrac{217 \pi ^2}{108}-16, 
		\end{alignat}
	\end{subequations}
	where $G=\text{Cl}_2(\frac{\pi}{3})\approx 1.0149$ denotes the Gieseking constant. The complete mass dependence
	separately for $C_{eB}$ and $C_{eW}$ is given in App.~\ref{app:complete-result} and provided as ancillary files.
	The above results include also the field normalisation contribution discussed in Sec.~\ref{sec:field-norm} together with the Higgs mass parameter
	terms from Sec.~\ref{sec:Higgs-mass}.
	
	As a final note, since the calculation was performed in naive dimensional regularization, 
	if the gauge boson attaches to the fermion loop in the last diagram in Fig.~\ref{fig:Ceg-2loop}, the resulting amplitude
	contains problematic traces of the form $\Tr{\gamma^\mu\gamma^\nu\gamma^\alpha\gamma^\beta\gamma^5}$. 
	However, since the amplitudes are expanded to first order in the external momenta two of the indices must be contracted with
	loop momenta which reduce either to a contraction with the metric tensor or directly to 0 after tensor reduction. In both cases the offending
	terms drop out identically. This was checked by computing the diagrams in both the "NDR" and "NDR-Discard" scheme of \texttt{FeynCalc},
	where the first keeps the problematic traces unevaluated in the final result while the second immediately drops them.

\section{Observables at NLO}\label{sec:resummed-obs}
\subsection{SMEFT renormalisation group equations}
Besides the significantly simplified structure of the two-loop result compared to a full mass-basis calculation,
the EFT framework also provides a direct way to obtain the leading SM corrections and resum the associated large logarithms
via the renormalisation group (RG) running in the EFT.
Here we specifically take into account the logarithmically enhanced gauge- and $y_t$ corrections using the SMEFT renormalisation group equations given in \cite{Jenkins:2013zja,Jenkins:2013wua,Alonso:2013hga}.
Neglecting the lepton Yukawa couplings and the dipole mixing in $C_{e\Phi}$, they reduce to
\begin{subequations}
	\begin{align}
		\dot{C}_{\underset{ji}{e\Phi}} &\simeq \Big[3N_c y_t^2 - \tfrac{21}{4} g_Y^2 - \tfrac{27}{4} g_W^2 + 24\lambda_\Phi \Big] C_{\underset{ji}{e\Phi}} \\
		\dot{C}_{\underset{ji}{eB}} &\simeq \Big[N_c y_t^2 + \tfrac{151}{12} g_Y^2 - \tfrac{9}{4}g_W^2\Big] C_{\underset{ji}{eB}} - \tfrac{3}{2} g_W g_Y C_{\underset{ji}{eW}}\\
		\dot{C}_{\underset{ji}{eW}} &\simeq \Big[N_c y_t^2 + \tfrac{1}{4}g_Y^2-\tfrac{11}{12}g_W^2\Big] C_{\underset{ji}{eW}} \, - \tfrac{1}{2} g_W g_Y C_{\underset{ji}{eB}}
	\end{align}
\end{subequations}
where $\dot{C}\equiv 16\pi^2 \bar{\mu} \frac{dC}{d\bar\mu}$. The gauge, top Yukawa and quartic coupling run according to the usual SM beta functions, given at one-loop by
\begin{subequations}
	\begin{align}
		\dot{g}_Y &= \tfrac{41}{6} g_Y^3, \qquad
		\dot{g}_W = -\tfrac{19}{6} g_W^3, \qquad
		\dot{g}_s = -7g_s^3, \\
		\dot{y}_t &= - y_t\Big[\tfrac{17}{12}g_Y^2 + \tfrac{9}{4}g_W^2 + 8 g_s^2 - \tfrac{9}{2} y_t^2\Big], \\
		\dot{\lambda}_\Phi &=4 N_c \lambda_\Phi y_t^2 - 2 N_c y_t^4 + \tfrac{3}{8} \big(g_Y^4 + 2g_Y^2 g_W^2 + 3g_W^4\big) - 3\lambda_\Phi\big(g_Y^2+3g_W^2+8\lambda_\Phi\big)
	\end{align}
\end{subequations}
where we neglect the mass suppressed WC contributions, that would affect the results only at dim-8. We also
neglect the threshold corrections to $g_Y$ and $g_W$ at the matching scale, which would affect the two-loop result
at $\O(\lambda_\text{VLL}^3 g_\text{SM}^2)$.

For the numerical evaluation below we solve the RGEs using the \texttt{Mathematica} function \texttt{NDSolve} with initial values
for the SM couplings at $\bar\mu_\text{EW}=M_Z$ (taken from Ref.~\cite{ParticleDataGroup:2026mpi}) and for the Wilson coefficients
at $\bar\mu_\Lambda=M_L$ from the matching conditions derived above.

\subsection{RG improved observables}

\begin{figure}[t]
	\begin{subfigure}{.49\textwidth}
		\includegraphics[width=.43\textwidth]{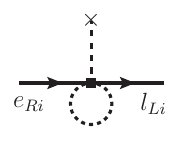}
		\includegraphics[width=.43\textwidth]{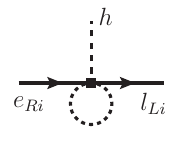}
		\caption{correction to $m_\ell$ and $\lambda_\ell$}
		\label{fig:SMEFT-1l-dim6-mass+Higgs}
	\end{subfigure}\hfill
	\begin{subfigure}{.49\textwidth}
		\includegraphics[width=.9\textwidth]{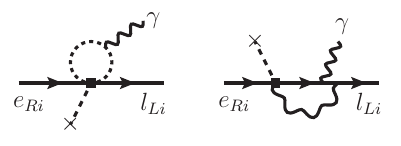}
		\caption{correction to $\Delta a_\ell$ and $d_\ell$}
		\label{fig:SMEFT-1l-dim6-dipole}
	\end{subfigure}
	\caption{One-loop SMEFT dim-6 corrections to \textbf{(a)} the lepton mass and Higgs coupling and
		\textbf{(b)} the lepton dipole moments. The scalar loop corresponds to a contraction of $\Phi^\dagger\Phi$
		from the effective operator which contains both the physical neutral component $h$ as well as the charged and neutral Goldstone
		bosons.}
\end{figure}

For simplicity we restrict the phenomenological discussion to the case where the VLL couple exclusively to a single generation of SM leptons. In that case all coupling matrices become diagonal and large flavour violating effects are avoided\footnote{For discussions of the phenomenology in case of flavour violating couplings see e.g. Refs.~\cite{Goudelis:2011un,Ishiwata:2013gma,Poh:2017tfo,Daberstiel:2026wxe}}. 
After evolving the Wilson coefficients to the EW scale $\bar\mu_\text{EW}=M_Z$, the LL resummed lepton mass and Higgs coupling matrices are given by
\begin{subequations}
	\begin{align}
		m_{\ell_i} &= y_{\ell_i} v + C_{\underset{ii}{e\Phi}} v^3 + \Delta m_{\ell_i}	\\
		\lambda_{\ell_i} &= y_{\ell_i} + 3 C_{\underset{ii}{e\Phi}} v^2 + \Delta \lambda_{\ell_i}
	\end{align}
\end{subequations}
where $\Delta m_{\ell_i}$ and $\Delta \lambda_{\ell_i}$ denote the SMEFT one-loop corrections 
which include both the dim-4 SM contribution (that are again of order $\O(\lambda_\text{VLL}^3 g_\text{SM}^2)$ and neglected here) 
as well as the dim-6 contribution from $C_{e\Phi}$ shown in Fig.~\ref{fig:SMEFT-1l-dim6-mass+Higgs}.
While the latter in principle contribute at the relevant $\O(\lambda_\text{VLL}^5)$, they drop out in the correlation between
the lepton mass and Higgs coupling, since here the Higgs correction is not enhanced by a combinatorial factor.
Therefore $\lambda_{\ell_i}$ including the leading NLO matching corrections is again given by Eq.~\eqref{eq:lepton-Higgs-LO}, however,
now with $C_{e\Phi}$ evaluated at $\bar\mu_\text{EW}=M_Z$ in order to resum the large logarithms from the SM gauge- and $y_t$ corrections.

Similarly, the one-loop dim-6 corrections to the dipole moments shown in Fig.~\ref{fig:SMEFT-1l-dim6-dipole} either vanish (left)
or contribute at $\O(\lambda_\text{VLL}^3g_\text{SM}^2)$ without logarithmic enhancement (right). In addition to the EW logs resummed by the
SMEFT RGE, we also include the universal leading QED logarithm from the LEFT running to the lepton mass scale 
\cite{Czarnecki:1995sz,Degrassi:1998es,Jenkins:2017dyc}, such that
\begin{subequations}
	\begin{align}
		\Delta a_{\ell_i} &= -\frac{4 m_{\ell_i} v}{e}\Re{C_{\underset{ii}{e\gamma}}(M_Z)} \Big[ 1 - \tfrac{4\alpha}{\pi} \ln(\tfrac{M_Z}{m_{\ell_i}})  \Big], \\
		d_{\ell_i} &= 2v\,\Im{C_{\underset{ii}{e\gamma}}(M_Z)} \Big[ 1 - \tfrac{4\alpha}{\pi} \ln(\tfrac{M_Z}{m_{\ell_i}})  \Big].
	\end{align}
\end{subequations}

\section{Numerical evaluation}\label{sec:numeric}
\begin{table}
	\centering
	\begin{tabular}{|c|cc|cc|cc|}
		\hline
		$\ell_i$ & \multicolumn{2}{c|}{$e$} & \multicolumn{2}{c|}{$\mu$} & \multicolumn{2}{c|}{$\tau$} \\\hline\hline
		$R_{\ell_i\ell_i}$ 	& $\lesssim 55000$ 				& \cite{CMS:2022urr} & $1.21\pm 0.35$ 				 & \cite{ParticleDataGroup:2024cfk} &	$0.91\pm 0.09$ 		 & \cite{ParticleDataGroup:2026mpi}\\
		$\Delta a_{\ell_i}$ & $\lesssim 10^{-12}$ 			& \footnotemark		
		 & $(3.8\pm6.3)\times 10^{-10}$ & \cite{Aliberti:2025beg,Muong-2:2025xyk} & -- 					 & \\
		$|d_{\ell_i}|$ 		& $< 4.1\times 10^{-30} e$ cm 	& \cite{Roussy:2022cmp} & $<1.9\times 10^{-19} e$ cm   & \cite{Muong-2:2008ebm} & $\lesssim 10^{-17}e$ cm & \cite{Belle:2021ybo} \\\hline
	\end{tabular}
	\caption{Experimental bounds on the lepton--Higgs coupling and dipole moments used in this paper.}
	\label{tab:exp-bounds}
\end{table}
\footnotetext{Based on the two different theory results using
	the current measurements of $\alpha$ \cite{Parker:2018vye,Morel:2020dww} which differ by more than $5\sigma$. See also discussion in \cite{Athron:2025ets}.}

We are now ready to evaluate the impact of the next-to-leading order corrections to the lepton--Higgs coupling and dipole moments
on the correlation discussed in Sec.~\ref{sec:LO-correlation}. We focus first on the most interesting muon-specific case where
the current bounds on $R_{\mu\mu}$ and $\Delta a_\mu$ as well as the expected future sensitivity $6\times 10^{-23}e$ cm for $d_\mu$ \cite{Sakurai:2022tbk}
are all of the right order of magnitude to test the correlation predicted by the vector-like leptons (see Tab.~\ref{tab:exp-bounds}).

\begin{figure}[t]
	\centering
	\begin{subfigure}{.49\textwidth}
		\sbox{\figbox}{\includegraphics[width=.5\textwidth]{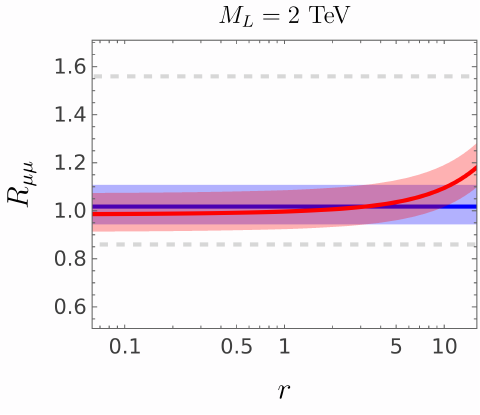}}
		\usebox{\figbox}
		\includegraphics[height=\ht\figbox,clip,trim=42 0 0 0]{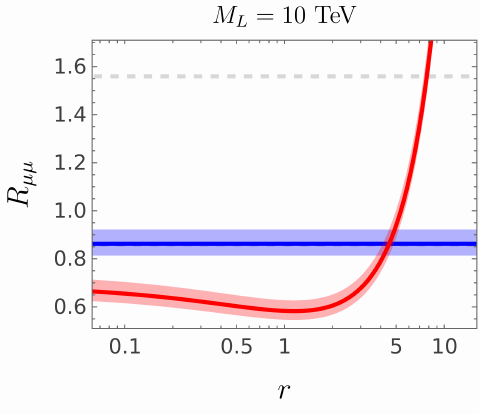}
		\caption{}
	\end{subfigure}
	\hfill
	\begin{subfigure}{.49\textwidth}
		\sbox{\figbox}{\includegraphics[width=.5\textwidth]{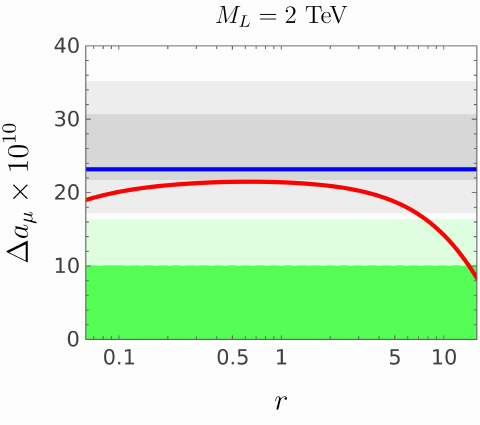}}
		\usebox{\figbox}
		\includegraphics[height=\ht\figbox,clip,trim=40 0 0 0]{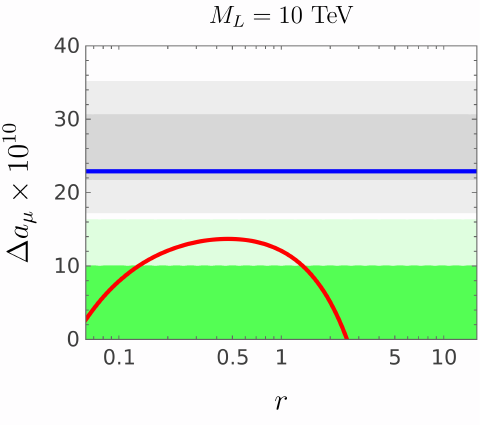}
		\caption{}
	\end{subfigure}
	\caption{Comparison of leading order (blue) and next-to-leading order (red) contributions to $R_{\mu\mu}$ \textbf{(a)}
		and $\Delta a_\mu$ \textbf{(b)} as a function of $r$ for $M_L=2$ TeV (left) or 10 TeV (right) 
		and fixed couplings $\lambda^{L,E}_\mu = 0.03\frac{M_{L,E}}{v}$, $\lambar=-0.75$ and $\lambda=0$.
		The band indicates the uncertainty from matching scale variation. The gray dashed lines show the current limit on $R_{\mu\mu}$,
		while the green (gray) bands show the 1 and 2 $\sigma$ limits from $\Delta a_\mu^\text{WP25}$ ($\Delta a_\mu^\text{WP20}$).
		Both the LO and NLO results are RG improved resulting in the small shift of the LO result for different $M_L$.
	}
	\label{fig:uncertainty}
\end{figure}

Fig.~\ref{fig:uncertainty} shows a typical (CP-conserving) scenario as a function of $r$ with $\lambda_\mu^{L,E}$ fixed close to the EWPO limit \cite{Daberstiel:2026wxe}, 
where the leading order prediction (blue) for $R_{\mu\mu}$ agrees with the current experimental bound and $\Delta a_\mu$ with the WP20 deviation
\cite{Aoyama:2020ynm,Athron:2025ets}. 
The red curves show the corresponding NLO result. Since here the VLL-SM couplings are scaled with the VLL mass (in agreement with the EWPO bounds) 
the $r$ dependence drops out exactly in the LO result. 
However, since overall the NLO corrections have the same mass suppression as the LO result, their effect
is increased with an increasing VLL mass scale as long as the couplings saturate the EWPO bounds. 
Furthermore a strong $r$ dependence is introduced via the loop functions (see App.\ref{app:complete-result}).
In addition the plots also show a rough uncertainty estimate based on varying the matching scale from $M_L/2$ to $2M_L$, which simulates the missing 
NLO gauge and $y_t$ corrections. As expected, for both $R_{\mu\mu}$ and the muon dipole moments this uncertainty is typically significantly smaller
than the pure VLL corrections and also the current experimental sensitivity, justifying our approximation made in the calculation of the NLO matching conditions above.

\begin{figure}[t]
	\centering
	\sbox{\figbox}{\includegraphics[width=.26\textwidth]{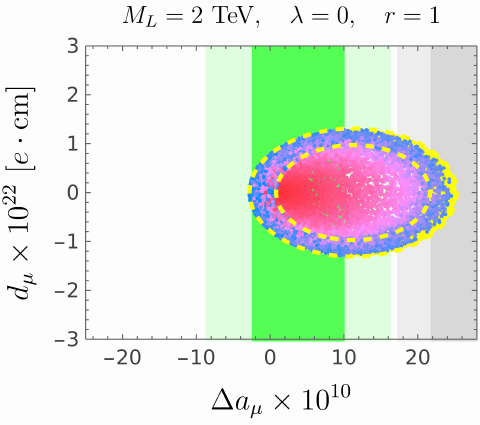}}
	\usebox{\figbox}
	\includegraphics[height=\ht\figbox,clip,trim=40 0 0 0]{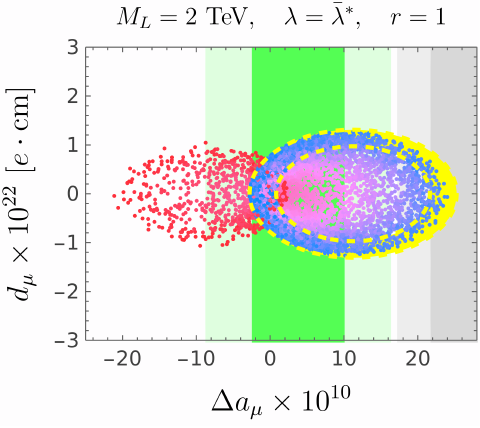}
	\includegraphics[height=\ht\figbox,clip,trim=40 0 0 0]{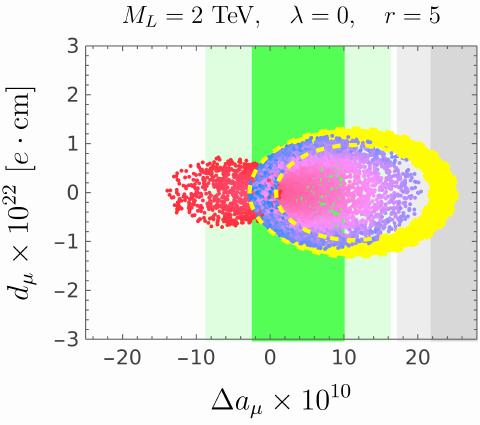}
	\includegraphics[height=\ht\figbox,clip,trim=40 0 0 0]{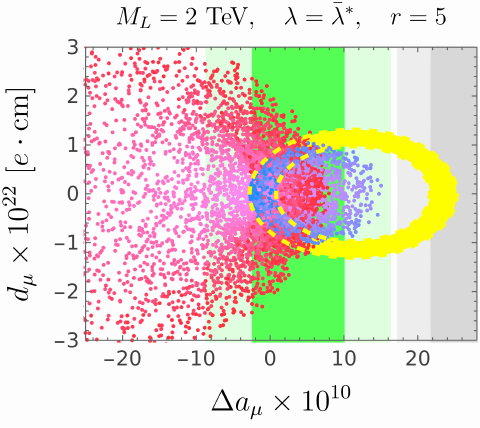}
	\raisebox{.5cm}{\includegraphics[height=.8\ht\figbox]{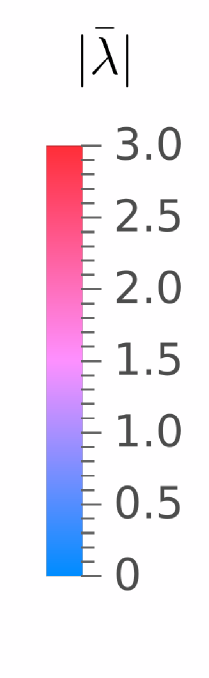}} \\[.2cm]
	\sbox{\figbox}{\includegraphics[width=.26\textwidth]{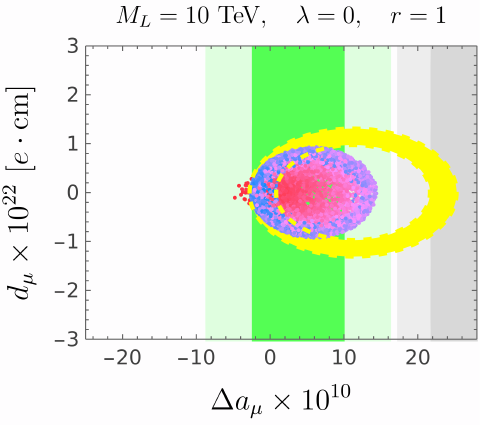}}
	\usebox{\figbox}
	\includegraphics[height=\ht\figbox,clip,trim=40 0 0 0]{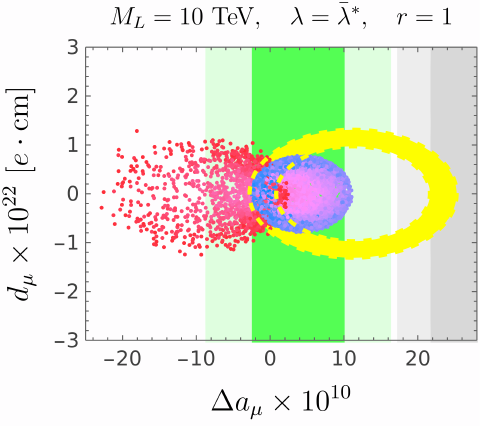}
	\includegraphics[height=\ht\figbox,clip,trim=40 0 0 0]{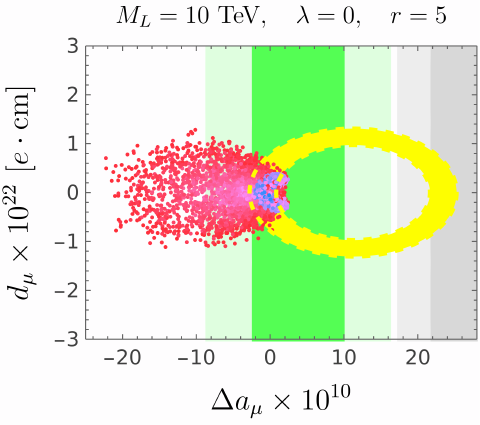}
	\includegraphics[height=\ht\figbox,clip,trim=40 0 0 0]{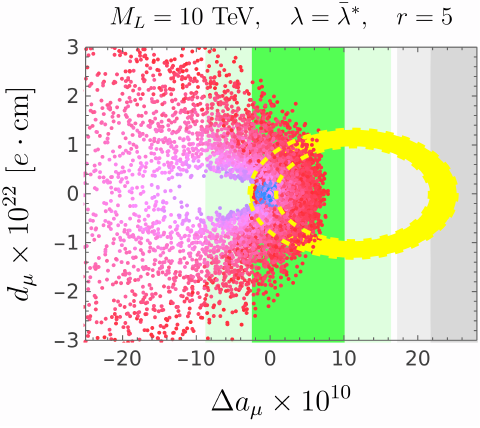}
	\raisebox{.5cm}{\includegraphics[height=.8\ht\figbox]{Figures/Legend.pdf}}
	\caption{Scatter plots of the random scan results for the NLO and RG improved observables. The couplings were sampled uniformly
		in the range $|\lambar|\in[0,3]$ (indicated by the point colour) with $\arg\lambar\in[-\pi,\pi]$ and $\lambda^{L,E}_\mu\in [0,0.5]$ (for $M_L=2$ TeV) and
		$[0,2]$ (for $M_L=10$ TeV) and the points are selected to fulfil the $R_{\mu\mu}$ bound. 
		The yellow contour shows the LO correlation (without RG improvement), while the green (gray) contours again
		show the 1 and 2 $\sigma$ limits from $\Delta a_\mu^\text{WP25}$ ($\Delta a_\mu^\text{WP20}$) (see also Ref.~\cite{Athron:2025ets}). }
	\label{fig:muon-ellipse-NLO}
\end{figure}

The full impact on the dipole ellipse is illustrated in Fig.~\ref{fig:muon-ellipse-NLO}. The scatter plots show points from random scans
in different sections of parameter space. The NLO corrections generally lead to a contraction of the ellipse compared to the LO result.
The strength of the contraction increases if either the mass splitting, the magnitude of (one of the) couplings or the VLL mass scale increases.
In fact, for $\lambar\gtrsim 2$
the points start to form a mirrored version of the ellipse\footnote{Note that VLL Yukawa couplings significantly above $1$ destabilize the EW vacuum and
	are therefore strongly constrained in pure VLL extensions \cite{Gopalakrishna:2018uxn,Cingiloglu:2024vdh,Adhikary:2024esf}.
	In addition the fact that $\kappa$ deviates strongly from the LO value for such couplings also signals a breakdown
	of perturbation theory, at least for the dipole coefficients, and thus provides another reason to restrict phenomenological 
	analyses to the region $\lambda_\text{VLL}\lesssim 1$.}  which now extends to negative values of $\Delta a_\mu$. The reason is that the $\O(1\ell)$
terms in Eq.~\eqref{eq:kappa-NLO} become sufficiently large to flip the sign of $\kappa$ (defined by Eq.~\eqref{eq:def-kappa}
in terms of the NLO Wilson coefficients at the matching scale). Note that for the scenarios shown ($\lambda=0$ or $\lambda=\lambar^*$)
$\kappa$ remains real, while in general the NLO corrections will also induce a phase leading to a slight shift of the centre of the ellipse.

\begin{figure}[t]
	\centering
	\sbox{\figbox}{\includegraphics[width=.25\textwidth]{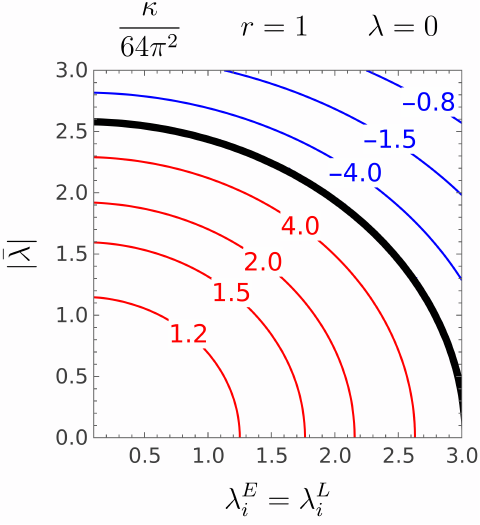}}
	\usebox{\figbox}
	\includegraphics[height=\ht\figbox,clip,trim=43 0 0 0]{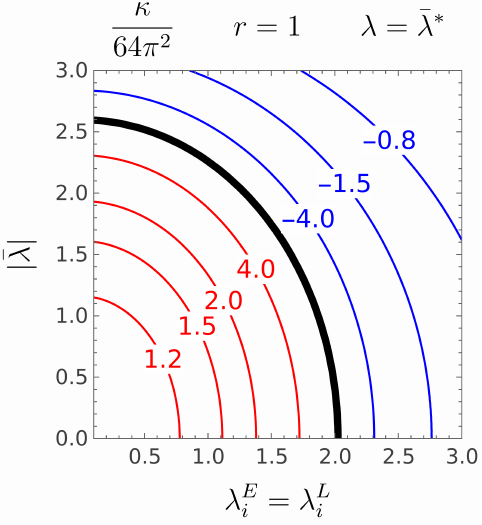} \hspace{.3cm}
	\sbox{\figbox}{\includegraphics[width=.25\textwidth]{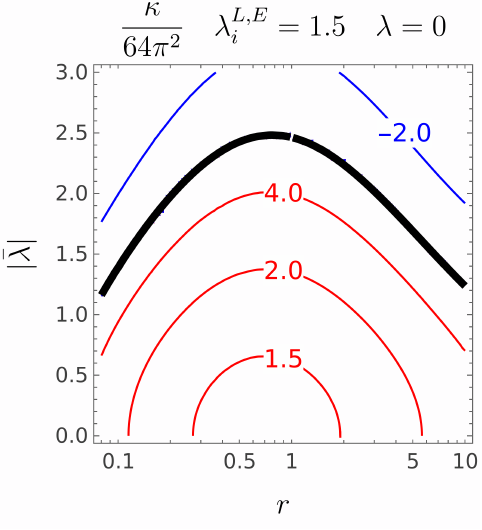}}
	\usebox{\figbox}
	\includegraphics[height=\ht\figbox,clip,trim=43 0 0 0]{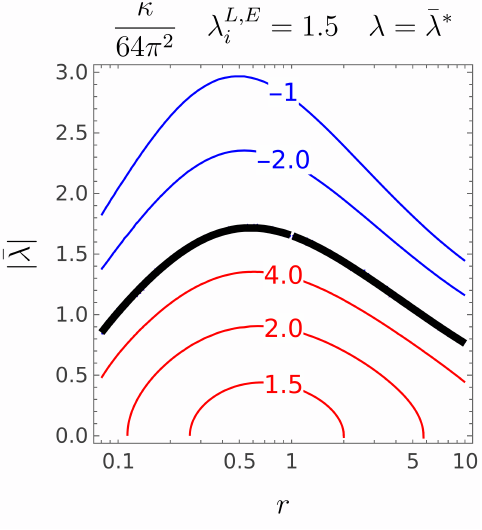}
	\caption{Contour plots of $\kappa$ in the $\lambda^{E,L}$--$\lambar$ (left) and $r$--$\lambar$ (right) plane. 
		The red (blue) contours, labelled with the respective values, show the regions where $\kappa$ is positive (negative) while 
		the black line is the singularity corresponding to $C_{e\gamma}=0$. Note that $\kappa$ is both independent of the absolute
		VLL mass scale as well as which flavour-specific scenario is realized (i.e. whether $i=e,\mu$ or $\tau$).}
	\label{fig:kappa-contours}
\end{figure}

Fig.~\ref{fig:kappa-contours} shows contours of $\kappa$ as a function of $\lambar$ and $\lambda^{L,E}$ or $r$. For moderate values of the couplings and $r\simeq 1$,
$\kappa$ is positive and close to the LO value. If $r$ deviates from 1, $\kappa$ remains positive, but can differ significantly from the LO result.
For increasing couplings $\kappa$ increases very quickly up until $C_{e\gamma}=0$ where the coefficient becomes singular. For larger values of the couplings
$\kappa$ then turns negative resulting in the mirrored ellipses in Fig.~\ref{fig:muon-ellipse-NLO}.

\begin{figure}[t]
	\centering
	\sbox{\figbox}{\includegraphics[width=.25\textwidth]{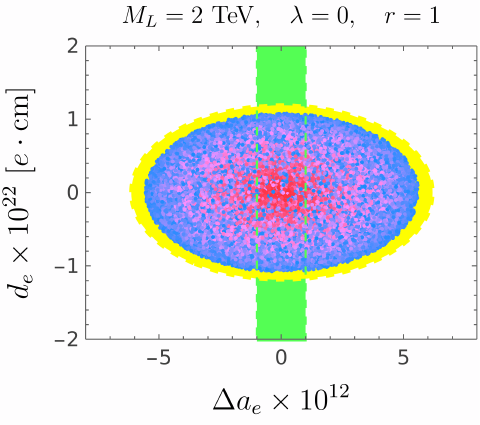}}
	\usebox{\figbox}
	\includegraphics[height=\ht\figbox,clip,trim=40 0 0 0]{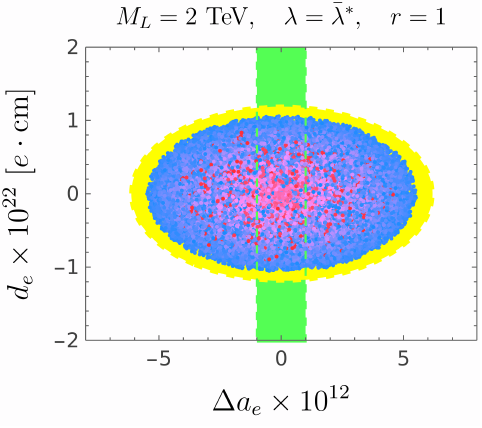}\hfill
	\includegraphics[height=\ht\figbox,clip,trim=0 0 0 0]{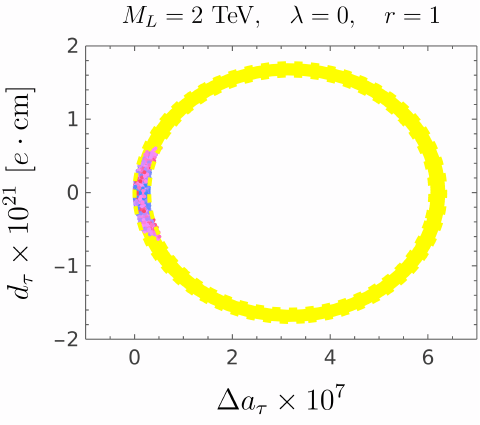}
	\includegraphics[height=\ht\figbox,clip,trim=40 0 0 0]{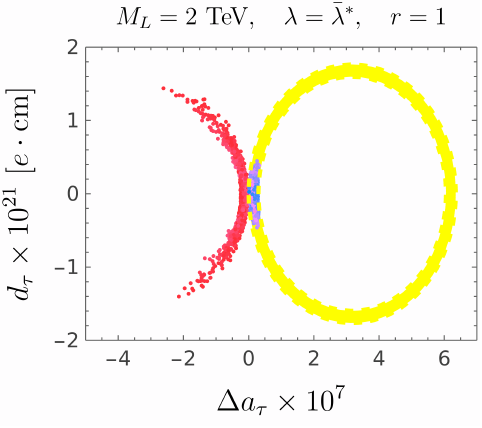}
	\raisebox{.5cm}{\includegraphics[height=.8\ht\figbox]{Figures/Legend.pdf}} 
	\caption{Scatter plots of the electron and tau ellipses. The parameter ranges are the same as in Fig.~\ref{fig:muon-ellipse-NLO}
		and the points are selected to fulfill $R_{ee}$ and $R_{\tau\tau}$ respectively. The yellow contour again shows the LO correlation. 
		The green band shows the (rough) bound $|\Delta a_e|<10^{-12}$. 
		Note that the current electron EDM bound essentially constrains the points to the x-axis,
		while the current bounds on the tau dipole moments allow the entire region shown in the plot. }
	\label{fig:e+tau-ellipse}
\end{figure}

Finally, we also briefly look at the effect on the electron and tau correlation. In both cases at least one of the current bounds 
is either significantly stronger or weaker (see Tab.~\ref{tab:exp-bounds}) than the scales relevant for the dipole ellipse. In the electron-specific
scenario shown in Fig.~\ref{fig:e+tau-ellipse} (left), 
the CP-violating EDM is already strongly constrained, essentially enforcing CP-conservation. This either requires extremely small couplings,
in which case also the predicted magnetic moment and $R_{ee}$ are vanishingly small, or a fine-tuned alignment of the CP-violating phases
(e.g. $\arg\lambar\lesssim 10^{-9}$). At the same time, the electron--Higgs coupling is currently very weakly constrained,
as a result the ellipse is basically centred at $(0,0)$ with the $\Delta a_e$ and $d_e$ bounds restricting the allowed values to an extremely narrow 
rectangular region around this centre. Unlike for the muon where the NLO corrections significantly deform the ellipse pulling previously
excluded regions back into the range compatible with the dipole bounds, the electron ellipse is only slightly contracted, resulting in
a minimal impact on the allowed parameter space.

Lastly, in the $\tau$-specific case, shown in Fig.~\ref{fig:e+tau-ellipse} (right), 
the measurement of $R_{\tau\tau}$ restricts the predicted dipole moments to values significantly below the
current bounds, though potentially in reach of future sensitivities \cite{Crivellin:2021spu,USBelleIIGroup:2022qro}.
However, the chiral enhancement required to reach the right-hand side of the ellipse is at least a factor of 10 larger than in case of the muon, due to 
the larger $\tau$ mass. At the same time the EWPO bounds on $\lambda^{E,L}_\tau$ are similar \cite{Daberstiel:2026wxe}, 
such that values of $\lambar\gtrsim 5$ are necessary to reach these points.
As a result the points in our scan never populate the entire ellipse, but only a small part close to $(0,0)$. Furthermore, since the behaviour
of $\kappa$ is independent of the flavour-specific scenario, at NLO the right side of the ellipse will never be populated since the large values of $\lambar$
required will also flip the sign of $\kappa$ resulting in the mirrored ellipse.

\section{Summary and conclusions}\label{sec:conclusions}

In this paper we have considered an extension of the SM by a 4th generation of vector-like leptons.
This model is part of a special class of scenarios where the contribution to the lepton--Higgs coupling and dipole moments
is strongly correlated \cite{Kannike:2011ng,Dermisek:2022aec}.
In the CP-conserving case the leading-order correlation leaves two options: 
either $\lambda_{\ell_i}\approx\lambda_{\ell_i}^\text{SM}$ and $\Delta a_{\ell_i}\approx0$, or 
$\lambda_{\ell_i}\approx-\lambda_{\ell_i}^\text{SM}$ and $\Delta a_{\ell_i}$ is fixed to a specific non-zero value. 
With CP-violation the correlation becomes an ellipse in the $\Delta a_{\ell_i}$--$d_{\ell_i}$ plane,
such that intermediate values of $\Delta a_{\ell_i}$ are possible, but require a non-zero $d_{\ell_i}$, which
will be further probed by future experiments like the muon EDM \cite{Sakurai:2022tbk} and measurements of the lepton--Higgs couplings at the HL-LHC \cite{Cepeda:2019klc}.

In this context, the $L\oplus E$ model was in particular motivated by the fact that the opposite-sign case predicts 
$\Delta a_\mu\approx2.3\times10^{-9}$ at leading order. This matched the deviation based on the data-driven HVP (WP20~\cite{Aoyama:2020ynm}), but is now excluded by the updated SM prediction (WP25~\cite{Aliberti:2025beg}). 

While the underlying VLL couplings drop out in the LO correlation, large underlying couplings (particularly for heavy VLL)
are nonetheless required in order to reach the opposite sign limit. This raises the question to what extent the LO predictions are reliable.
Ref.~\cite{Mohling:2024qvk} showed that the NLO corrections to the lepton--Higgs coupling are sizeable and can pull $\Delta a_\mu$ back towards the WP25 range. 
With this work we provide the missing part to complete the correlation at NLO; the two-loop corrections to the dipole moments.

To compute these corrections we have pursued an EFT approach that both simplifies the actual two-loop calculations and allow us to resum large logarithmic corrections
via the EFT renormalisation group equations, significantly improving the accuracy compared to a fixed order calculation.
We have computed the SMEFT matching conditions at NLO in the VLL Yukawa couplings for arbitrary masses of the VLL. 
Beyond the genuine two-loop diagrams, two contributions have to be taken into account. 
The first is the finite field redefinition required to restore canonically normalized EFT fields. 
The second is the threshold correction to the Higgs mass parameter.
Because $\Delta\mu^2$ is loop suppressed but proportional to the heavy VLL masses, 
it turns the formally dimension-8 one-loop contribution of order $\O(\mu^2_{\rm UV}/M^4_\text{VLL})$ into a dimension-6 contribution at NLO. 

We checked the cancellation of all UV and IR divergences and of the spurious singularities in the mass ratio was and also compared
the one-loop results with \texttt{Matchete}, which agree.
The complete mass dependence of the matching conditions is listed in App.~\ref{app:complete-result} and also provided as ancillary files. 

To obtain predictions for the lepton mass, the Higgs coupling and the dipole moments we have combined the matching conditions with the 
leading-logarithmic SMEFT and LEFT running to obtain the RG-improved observables at NLO.
The NLO corrections can become numerical large if the VLL couplings become $\O(1)$ or mass ratio deviates from $1$. 
Generically, the effect of the NLO corrections is a contraction of the dipole ellipse pulling the points towards smaller values of $\Delta a_\mu$.
For the muon this reinforces the trend found in Ref.~\cite{Mohling:2024qvk}: for heavy VLL and $M_E\neq M_L$ 
the opposite-sign scenario can again become compatible with WP25, while for $M_L\sim2$\,TeV and $r\simeq1$ it stays close to the 
leading-order value (see Fig.~\ref{fig:muon-ellipse-NLO}). Note that for sufficiently large deviations from $r=1$ it might become necessary
to integrate out the VLL sequentially in order to resum the $\ln(r)$ terms, similar to the logarithmic SM corrections.

However, in contrast to the fixed order mass basis calculation, in the RG improved EFT
approach the NLO corrections to $R_{\ell_i\ell_i}$ actually remain rather moderate. This is because the matching avoids 
large logarithmic term associated with the VLL couplings, which where driving the large corrections found in Ref.~\cite{Mohling:2024qvk}. 
Instead, the shift in the correlation is mainly a result of the NLO corrections to the dipole moments, which are thus essential to obtain
a reliable prediction in the VLL model.

For couplings $\gtrsim2$ the NLO terms become comparable to the leading order and
$C_{e\gamma}$ can even change its sign compared to LO, such that the correlation turns into a mirrored ellipse at negative $\Delta a_\ell$. 
However, this region is strongly constrained by vacuum stability \cite{Gopalakrishna:2018uxn,Cingiloglu:2024vdh,Adhikary:2024esf}
and the relative size of the NLO correction signals a breakdown of the perturbative expansion rather than a genuine physical effect.

The behaviour for the three lepton generations differs markedly because of the available data. 
For the muon, the current bounds on $R_{\mu\mu}$ and $\Delta a_\mu$ as well as the projected sensitivity of $d_\mu$ are all of the right size to test the correlation, so the NLO corrections are directly relevant. 
For the electron, the bounds on $\Delta a_e$ and $d_e$ are far stronger than those on $R_{ee}$. 
The electron EDM essentially enforces CP-conservation, and the NLO contraction of the ellipse has only a small impact on the allowed parameter space. 
For the $\tau$, the measurement of $R_{\tau\tau}$ restricts the dipole moments to values well below current bounds but 
potentially within reach of future experiments. However, reaching the far side of the ellipse at LO requires couplings $\gtrsim 5$, which in turn
results in the flipped sign of $C_{e\gamma}$ at NLO, such that this part is never populated.

Several future extensions of this work are possible. The results can also be applied to flavour-violating observables to evaluate the implications
for phenomenological studies like Ref.~\cite{Daberstiel:2026wxe}.
Another improvement of the results would be an inclusion of the gauge and top-Yukawa contributions to the two-loop matching which 
would remove the dominant remaining uncertainty. Furthermore, the calculation could be extended to VLL scenarios with a larger Higgs sector, 
where the dipole--Higgs correlation takes a similar form \cite{Chun:2020uzw,Raju:2022zlv,Dermisek:2023tgq}.

\section*{Acknowledgments}

I want to thank Dominik Stöckinger for his comments and encouragement to publish these results
as well as Paul Kühler for our discussions on the $\gamma^5$-related issues.
Part of this manuscript and the code to set up the parameter scans were created with the help
of Claude Opus 5.5. The Feynman diagram figures were created using \texttt{Axodraw} \cite{Collins:2016aya}.
This work was financially supported by the German Research Foundation
(DFG) under grant number STO 876/7-2 and STO 876/10-1.

\appendix
\allowdisplaybreaks
\section{Complete mass dependence}\label{app:complete-result}
Here we collect the complete dependence on the mass ratio $r=M_E^2/M_L^2$ of the coefficients in Eqs.~\eqref{eq:CeH-1loop}
and \eqref{eq:Ceg-2loop}. These expressions are also provided as ancillary \texttt{Mathematica} files.

\subsection{$C_{e\Phi}$ coefficients}
	The full mass dependence of the coefficients in Eq.~\eqref{eq:CeH-1loop} is given by
	\begin{subequations}
		\begin{align}
			c_1 &= \tfrac{1}{8} \Big[28 L(M_E) - 17\Big], \\[.2cm]
			c_2 &= \tfrac{1}{2}\Big[8 L(M_L) - 5\Big], \\[.2cm]
			c_3 &= -\tfrac{7}{4}+\tfrac{3 r}{(r-1)^2}+3 \, L(M_L)+\left(\tfrac{3}{2}-\tfrac{3 (3 r-1)}{2 (r-1)^3}\right) \ln (r), \\[.2cm]
			c_4 &= \tfrac{1}{\sqrt{r}}\bigg[\tfrac{3}{2}+\tfrac{3 (3 r-1)}{2 (r-1)^2}+\left(\tfrac{-4 r^2+2r-1}{(r-1)^3}-1\right) \ln (r)\bigg], \\[.2cm]
			c_5 &= \sqrt{r}\bigg[2+\tfrac{4}{r}+\tfrac{3 (r+1)}{2 (r-1)^2}-\Big(\tfrac{8}{r}+4\Big)\, L(M_L)+\left(\tfrac{r^2-5 r+1}{(r-1)^3}-2\right) \ln (r)\bigg].
		\end{align}
	\end{subequations}

\subsection{$C_{eB}$ coefficients}
	$\Delta_{2\ell}C_{eB}$ has the same structure as $\Delta_{2\ell}C_{e\gamma}$ in Eq.~\eqref{eq:Ceg-2loop} except for $e\to g_Y$. The coefficients read
	\begin{subequations}
	\begin{align}
		\begin{split}
			&a_1= -\tfrac{480 r+137}{4}-\tfrac{2}{r}+\tfrac{\pi ^2}{6} \left( 96 r^2-97 r+44 + \tfrac{3 r^2-2 r-2}{r^3}\right) +3 (16 r+37) \, L(M_L) \\
			& +\left(\tfrac{-19 r^2+8 r+4}{r^3}-96 r-\tfrac{36}{r-1}-47\right) \tilde{I}(0,r)
				 +\left(\tfrac{2 \left(4 r^2-3 r-1\right)}{r^3}+\tfrac{3 (24 r-7)}{2 (4 r-1)^2}-\tfrac{3}{2}\right) \tilde{I}(r,r) \\
			& +\left(\tfrac{1}{2} (107-144 r)+\tfrac{4}{r}+\tfrac{3}{4 r-1}\right) \ln (r)
				 +\left(\tfrac{3 (24 r-7)}{4 (4 r-1)^2}-\tfrac{192 r+97}{4}-\tfrac{18}{r-1}-\tfrac{4}{r}\right) \ln ^2(r),
		\end{split}\\[.5cm]
		\begin{split}
			 &a_2= -169-\tfrac{12}{r}+2 r+\tfrac{\pi ^2}{6} \left(\tfrac{61 r-12}{r^2}+ 2 r^3+8 r^2-3 r-8\right) + \left(\tfrac{48}{r}+120\right) \, L(M_L) \\
			& +\left(\tfrac{12-25 r}{r^2}-4 r^2-20 r+\tfrac{24}{r-1}+7\right) \tilde{I}(0,r)
			 	+2\left( r^2+6 r+2 + \tfrac{3 (7 r-30)}{(r-4)^2} + \tfrac{3}{r}\right) \tilde{I}(r,1) \\
			& +\left(4 r+\tfrac{24}{r-4}-\tfrac{12}{r}+25\right) \ln (r) 
			 	+\left(\tfrac{8 r+47}{2}+\tfrac{6 (13 r-56)}{(r-4)^2}+\tfrac{12}{r-1}+\tfrac{6}{r}\right) \ln ^2(r),
		\end{split}\\[.5cm]
		\begin{split}
			 &a_3= \tfrac{36}{r}-\tfrac{192 r+347}{2}+\tfrac{2 (8 r+1)}{(r-1)^2}+ \tfrac{\pi^2}{6} \left(\tfrac{49 r-48}{r^2} -24 r^2-25 r-18\right) 
				+6 \left(8 r+25+\tfrac{8}{r}\right) \, L(M_L)\\
			& +\left(\tfrac{3 (13 r+12)}{r^2}-\tfrac{24 (5 r-3)}{(r-1)^2}+3 (40 r+19)\right) \tilde{I}(0,r) 
				+\left(\tfrac{10 (4 r-1)}{3 (r-1)^2}-\tfrac{4 (r-3)}{r^2}+\tfrac{784 r-163}{6 (4 r-1)^2}-\tfrac{3}{2}\right) \tilde{I}(r,r)\\
			& +\left(\tfrac{6}{r}-\tfrac{10 (5 r-2)}{3 (r-1)^2}-8 (12 r+19)-\tfrac{4 (256 r-1093)}{3 (r-4)^2}\right) \tilde{I}(r,1) \\
			& +\left(\tfrac{-76 r^2+127 r-69}{(r-1)^3}-48 r-\tfrac{184}{r-4}-\tfrac{11}{4 r-1}-\tfrac{60}{r}-139\right) \ln (r) \\
			& +\left(\tfrac{281-353 r}{3 (r-1)^2}-\tfrac{144 r+881}{4} +\tfrac{784 r-163}{12 (4 r-1)^2}-\tfrac{2 (955 r-4096)}{3 (r-4)^2}+\tfrac{30}{r}\right) \ln ^2(r),
		\end{split}\\[.5cm]
		\begin{split}
			 &a_4= \tfrac{36}{r}+\tfrac{18 r}{(r-1)^2}-\tfrac{3}{2} (64 r+113)+\pi ^2 \left(\tfrac{9 r-8}{r^2}+\tfrac{1}{2} \left(-8 r^2-8 r-5\right)\right) \\
			& +\left(\tfrac{9 (3 r+4)}{r^2}-\tfrac{24 (5 r-3)}{(r-1)^2}+3 (40 r+17)\right) \tilde{I}(0,r) 
				+\left(\tfrac{3 (r+4)}{r^2}+\tfrac{25 r-13}{3 (r-1)^2}+\tfrac{640 r-127}{6 (4 r-1)^2}-\tfrac{3}{2}\right) \tilde{I}(r,r)\\
			& +\left(\tfrac{17-29 r}{3 (r-1)^2}-3 (32 r+49)-\tfrac{2 (503 r-2150)}{3 (r-4)^2}+\tfrac{6}{r}\right) \tilde{I}(r,1) + 6\left(8 r+25+\tfrac{8}{r}\right) \, L(M_L) \\
			& -\left(\tfrac{9 \left(8 r^2-13 r+7\right)}{(r-1)^3} + 3 (16 r+45) + \tfrac{184}{r-4} + \tfrac{60}{r} + \tfrac{11}{4 r-1}\right) \ln (r) \\
			& +\left(\tfrac{640 r-127}{12 (4 r-1)^2}-\tfrac{9}{4} (16 r+97)-\tfrac{2 (175 r-139)}{3 (r-1)^2}-\tfrac{2 (937 r-4024)}{3 (r-4)^2}+\tfrac{30}{r}\right) \ln ^2(r),
		\end{split}\\[.5cm]
		\begin{split}
			\sqrt{r} \, &a_5 = \tfrac{9 (3 r-1)}{(r-1)^2}-3 (44 r-59)+\pi ^2 \left(-8 r^2-\tfrac{16}{r}+11\right) - 24 (r+4) \, L(M_L) \\
			& +\left(-\tfrac{48 (4 r-3)}{(r-1)^2}+48 (5 r-1)+\tfrac{72}{r}\right) \tilde{I}(0,r) 
				+\left(\tfrac{2 (23 r-8)}{3 (r-1)^2}+\tfrac{5 (80 r-17)}{6 (4 r-1)^2}+\tfrac{24}{r}+\tfrac{3}{2}\right) \tilde{I}(r,r)  \\
			& - \left(48 (4 r+3)+\tfrac{2 (28 r-13)}{3 (r-1)^2}+\tfrac{4 (427 r-1840)}{3 (r-4)^2}\right) \tilde{I}(r,1) \\
			& -\left(\tfrac{18 \left(5 r^2-8 r+4\right)}{(r-1)^3} + 3 (52 r+95) + \tfrac{352}{r-4} + \tfrac{5}{4 r-1}\right) \ln (r) \\
			& +\left(\tfrac{389-461 r}{3 (r-1)^2}+\tfrac{5 (80 r-17)}{12 (4 r-1)^2}-\tfrac{3}{4} (96 r+463)-\tfrac{16 (197 r-854)}{3 (r-4)^2}\right) \ln ^2(r),
		\end{split}
	\end{align}
	\end{subequations}
	
	\begin{subequations}
	\begin{align}
		\begin{split}
			\sqrt{r}\,&\tilde{a}_1 = 2+\tfrac{\pi ^2}{6} \left(\tfrac{3 r-4}{r^2}+4r\right)
				+\left(\tfrac{3}{2 (4 r-1)}+\tfrac{7}{2}\right) \ln (r) +\left(\tfrac{3 (16 r-5)}{8 (4 r-1)^2}+\tfrac{3}{r-1}-\tfrac{1}{8}\right) \ln ^2(r)\\
				&+\left(\tfrac{6}{r-1}-\tfrac{4 (r-2)}{r^2}-4\right) \tilde{I}(0,r) 
				+\left(\tfrac{r-4}{r^2}+\tfrac{3 (16 r-5)}{4 (4 r-1)^2}+\tfrac{15}{4}\right) \tilde{I}(r,r),
		\end{split}\\[.5cm]
		\begin{split}
			\sqrt{r}\,&\tilde{a}_2 = 2 (2 r-3)+\tfrac{\pi^2}{6} \left(4 r^3-3 r^2-4+\tfrac{6}{r}\right)\\
			& + \left(8 r+\tfrac{48 (2 r-9)}{(r-4)^2}+\tfrac{3}{r-1}+27\right) \ln ^2(r) + \left(2 (4 r+9)+\tfrac{48}{r-4}\right) \ln (r)\\
			&-2\left(4 r^2+r-2-\tfrac{3}{r-1}+\tfrac{3}{r}\right) \tilde{I}(0,r) + \left(4 r^2+5 r + \tfrac{6 (9 r-40)}{(r-4)^2}+12\right) \tilde{I}(r,1),
		\end{split}\\[.5cm]
		\begin{split}
			\sqrt{r}\,&\tilde{a}_3 = -6+\pi ^2 \left(\tfrac{r+1}{2}+\tfrac{1}{r}\right) +\left(-\tfrac{16 (5 r-23)}{3 (r-4)^2}+\tfrac{32 r-5}{24 (4 r-1)^2}+\tfrac{19}{3 (r-1)}-\tfrac{33}{8}\right) \ln ^2(r)\\
			&+\left(-\tfrac{6}{r-1}-\tfrac{1}{2 (4 r-1)}-\tfrac{16}{r-4}-\tfrac{9}{2}\right) \ln (r) +\left(-\tfrac{6}{r}+\tfrac{18}{r-1}+6\right) \tilde{I}(0,r) \\
			& +\left(\tfrac{5-17 r}{3 (r-1)^2}-\tfrac{2 (23 r-104)}{3 (r-4)^2}-9\right) \tilde{I}(r,1)
			+\left(\tfrac{13 r-1}{3 (r-1)^2}+\tfrac{32 r-5}{12 (4 r-1)^2}+\tfrac{15}{4}\right) \tilde{I}(r,r),
		\end{split}\\[.5cm]
		\begin{split}
			\sqrt{r}\,&\tilde{a}_4 = 3-224 r+\tfrac{25 r-7}{(r-1)^2}+ \tfrac{\pi ^2}{6} \left(35-2 r^2-47r+\tfrac{8}{r}\right) + 48 (r+1) \, L(M_L) \\
			& +\left(3 (68 r-13)-\tfrac{6 (17 r-9)}{(r-1)^2}-\tfrac{36}{r}\right) \tilde{I}(0,r)
				+\left(\tfrac{4 (10 r-7)}{3 (r-1)^2}+\tfrac{13 (32 r-5)}{12 (4 r-1)^2}+\tfrac{28}{r}+\tfrac{35}{4}\right) \tilde{I}(r,r) \\
			& -\left(\tfrac{4 (11 r-8)}{3 (r-1)^2}+2 (101 r+81)+\tfrac{2 (929 r-4016)}{3 (r-4)^2}\right) \tilde{I}(r,1) \\
			& +\left(\tfrac{152 r-93 r^2-77}{(r-1)^3}-\tfrac{11(32 r+37)}{2} -\tfrac{400}{r-4}-\tfrac{13}{2 (4 r-1)}\right) \ln (r) \\
			& +\left(\tfrac{13 (32 r-5)}{24 (4 r-1)^2}-\tfrac{800 r+3033}{8}-\tfrac{8 (427 r-1858)}{3 (r-4)^2}+\tfrac{493-637 r}{6 (r-1)^2}\right) \ln ^2(r),
		\end{split}\\[.5cm]
		\begin{split}
			&\tilde{a}_5 = -8-\tfrac{4}{r-1}+ \tfrac{\pi ^2}{6} \left(3-2 r+\tfrac{8}{r}\right)+\left(-\tfrac{12}{r}+\tfrac{18}{r-1}+12\right) \tilde{I}(0,r) \\
			& +\left(\tfrac{2-5 r}{3 (r-1)^2}-\tfrac{2 (23 r-104)}{3 (r-4)^2}-10\right) \tilde{I}(r,1)
				 +\left(\tfrac{4 r-1}{3 (r-1)^2}+\tfrac{4}{r}+\tfrac{2 (10 r-1)}{3 (4 r-1)^2}\right) \tilde{I}(r,r) \\
			& +\left(-\tfrac{2 (r-3)}{(r-1)^2}-\tfrac{16}{r-4}-\tfrac{2}{4 r-1}-8\right) \ln (r) 
				 +\left(-\tfrac{16 (5 r-23)}{3 (r-4)^2}+\tfrac{10 r-1}{3 (4 r-1)^2}+\tfrac{16}{3 (r-1)}-4\right) \ln ^2(r),
		\end{split}
	\end{align}
	\end{subequations}
	where the function $\tilde{I}(x,y)$ is given in Eq.~\eqref{eq:Itil}.
	Note that the apparent poles at $r=1,4$ and $\frac{1}{4}$ cancel after inserting the analytic expression for $\tilde{I}$
	and evaluating the respective limits.

\subsection{$C_{eW}$ coefficients}
	$\Delta_{2\ell}C_{eW}$ has the same structure as $\Delta_{2\ell}C_{e\gamma}$ in Eq.~\eqref{eq:Ceg-2loop} except for $e\to g_W$. The coefficients read
	\begin{subequations}
	\begin{align}
		\begin{split}
			&a_1 = \tfrac{27}{2}-\tfrac{2}{r}+ \tfrac{\pi ^2}{6} \left(\tfrac{3 r^2-6 r-2}{r^3}+ 20-9 r\right)\\
			& +\left(\tfrac{-11 r^2+16 r+4}{r^3}+\tfrac{12}{r-1}+9\right) \tilde{I}(0,r) 
				+\left(-\tfrac{2 (5 r+1)}{r^3}+\tfrac{3 (88 r-21)}{2 (4 r-1)^2}-\tfrac{9}{2}\right) \tilde{I}(r,r) \\
			&  +30 \, L(M_L) +\left(-\tfrac{3}{4 r-1}+\tfrac{4}{r}+15\right) \ln (r)+\left(\tfrac{3 (88 r-21)}{4 (4 r-1)^2}+\tfrac{6}{r-1}-\tfrac{4}{r}+\tfrac{9}{4}\right) \ln ^2(r),
		\end{split}\\[.5cm]
		\begin{split}
			&a_2 = -29+\tfrac{36}{r}+2 r+ \tfrac{\pi ^2}{6} \left(\tfrac{53 r-36}{r^2}+ 2 r^3+4 r^2-3 r-8\right) \\
			& +\left(-4 r^2+\tfrac{7 r+36}{r^2}-12 r+15\right) \tilde{I}(0,r)+\left(2 \left(r^2+4 r-2\right)+\tfrac{21}{r-4}+\tfrac{3}{r}\right) \tilde{I}(r,1) \\
			& +36 \, L(M_L)+\left(4 r-\tfrac{36}{r}+5\right) \ln (r)+\left(\tfrac{1}{2} (8 r+31)+\tfrac{42}{r-4}+\tfrac{18}{r}\right) \ln ^2(r),
		\end{split}\\[.5cm]
		\begin{split}
			&a_3 = 18+ \tfrac{\pi ^2}{6} \left(22-9 r+\tfrac{29}{r}\right)+\tfrac{36}{r}+\tfrac{2 (5 r+1)}{(r-1)^2} +24 \, L(M_L)
				+\left(\tfrac{12 (r+3)}{(r-1)^2}-\tfrac{3 (7 r+12)}{r^2}+9\right) \tilde{I}(0,r) \\
			& +\left(\tfrac{25}{3 (r-4)}-\tfrac{10 (r+2)}{3 (r-1)^2}+\tfrac{3}{r}\right) \tilde{I}(r,1)
			 	+\left(\tfrac{4 (13 r+9)}{r^2}+\tfrac{213-896 r}{2 (4 r-1)^2}+\tfrac{10}{(r-1)^2}-\tfrac{9}{2}\right) \tilde{I}(r,r) \\
			& +\left(\tfrac{4 \left(17 r^2-38 r+18\right)}{(r-1)^3}-\tfrac{36}{r}+\tfrac{11}{4 r-1}+12\right) \ln (r)
			 +\left(\tfrac{213-896 r}{4 (4 r-1)^2}+\tfrac{50}{3 (r-4)}+\tfrac{18}{r}+\tfrac{83-11 r}{3 (r-1)^2}+\tfrac{9}{4}\right) \ln ^2(r),
		\end{split}\\[.5cm]
		\begin{split}
			&a_4 = 18+\pi ^2 \left(\tfrac{1}{2} (7-4 r)+\tfrac{5}{r}\right)+\tfrac{36}{r}+\tfrac{12 r}{(r-1)^2} +24 \, L(M_L) 
			 +\left(\tfrac{12 (r+3)}{(r-1)^2}-\tfrac{3 (7 r+12)}{r^2}+15\right) \tilde{I}(0,r) \\
			& +\left(\tfrac{19}{3 (r-4)}-\tfrac{r+11}{3 (r-1)^2}+\tfrac{3}{r}-3\right) \tilde{I}(r,1)
				+\left(\tfrac{3 (17 r+12)}{r^2}+\tfrac{5-r}{(r-1)^2}-\tfrac{11 (80 r-19)}{2 (4 r-1)^2}-\tfrac{9}{2}\right) \tilde{I}(r,r) \\
			& +\left(\tfrac{2 \left(32 r^2-73 r+35\right)}{(r-1)^3}-\tfrac{36}{r}+\tfrac{11}{4 r-1}+12\right) \ln (r) 
				 +\left(\tfrac{38}{3 (r-4)}-\tfrac{8 (r-10)}{3 (r-1)^2}-\tfrac{11 (80 r-19)}{4 (4 r-1)^2}+\tfrac{18}{r}+\tfrac{9}{4}\right) \ln ^2(r),
		\end{split}\\[.5cm]
		\begin{split}
			\sqrt{r}\,&a_5 = \pi ^2 (5-2 r)+6 (2 r+17)+\tfrac{6 (3 r-1)}{(r-1)^2} -24 (r+2) \, L(M_L) +\left(\tfrac{24 (r+1)}{(r-1)^2}-\tfrac{72}{r}\right) \tilde{I}(0,r) \\
			& +\left(-\tfrac{2 (14 r+1)}{3 (r-1)^2}+\tfrac{100}{3 (r-4)}+12\right) \tilde{I}(r,1)
				 +\left(\tfrac{91-384 r}{2 (4 r-1)^2}+\tfrac{2 (3 r+2)}{(r-1)^2}+\tfrac{72}{r}-\tfrac{15}{2}\right) \tilde{I}(r,r) \\
			& +\left(\tfrac{4 \left(13 r^2-32 r+16\right)}{(r-1)^3}-3 (4 r+1)+\tfrac{5}{4 r-1}\right) \ln (r) 
				 +\left(\tfrac{91-384 r}{4 (4 r-1)^2}+\tfrac{7 r+65}{3 (r-1)^2}+\tfrac{200}{3 (r-4)}+\tfrac{33}{4}\right) \ln ^2(r).
		\end{split}
	\end{align}
	\end{subequations}

	\begin{subequations}
	\begin{align}
		\begin{split}
			\sqrt{r}\,&\tilde{a}_1 = 2 (2 r-1)+ \tfrac{\pi ^2}{6} \left(\tfrac{3 r-4}{r^2}-4\left(r^2-r\right)\right)
				+\left(\tfrac{16 r+13}{8}+\tfrac{3 (64 r-15)}{8 (4 r-1)^2}+\tfrac{3}{r-1}\right) \ln ^2(r)\\
			& +\left(\tfrac{8 r+9}{2}-\tfrac{3}{2 (4 r-1)}\right) \ln (r)+\left(\tfrac{8}{r^2}+4 r+\tfrac{6}{r-1}\right) \tilde{I}(0,r) 
				+\left(\tfrac{3 (64 r-15)}{4 (4 r-1)^2}-\tfrac{3 r+4}{r^2}+\tfrac{13}{4}\right) \tilde{I}(r,r), \\
		\end{split}\\[.5cm]
		\begin{split}
			\sqrt{r}\,&\tilde{a}_2 = -2+\tfrac{\pi ^2}{6} \left(4 r^3-3 r^2-4+\tfrac{2}{r}\right) +2 (2 r+1) \ln (r)+\left(6 r+\tfrac{60}{r-4}+\tfrac{3}{r-1}+17\right) \ln ^2(r)\\
			& -2\left(4 r^2-r-4-\tfrac{3}{r-1}+\tfrac{1}{r}\right) \tilde{I}(0,r)+\left(4 r^2+r+\tfrac{30}{r-4}+4\right) \tilde{I}(r,1),
		\end{split}\\[.5cm]
		\begin{split}
			\sqrt{r}\,&\tilde{a}_3 = 2 (2 r-1)+\tfrac{\pi ^2}{6} \left(\tfrac{2}{r}-4 r^2-r-1\right)+2\left(2 r+5+\tfrac{9}{r-1}-\tfrac{1}{r}\right) \tilde{I}(0,r) \\
			& +\left(\tfrac{1-13 r}{3 (r-1)^2}-\tfrac{14}{3 (r-4)}-5\right) \tilde{I}(r,1) +\left(\tfrac{11-48 r}{8 (4 r-1)^2}+\tfrac{1}{8} (16 r+13)-\tfrac{28}{3 (r-4)}+\tfrac{19}{3 (r-1)}\right) \ln ^2(r)\\
			& +\left(\tfrac{11-48 r}{4 (4 r-1)^2}+\tfrac{3 r+1}{(r-1)^2}+\tfrac{13}{4}\right) \tilde{I}(r,r) +\left(\tfrac{1}{2} (8 r+9)+\tfrac{2}{r-1}+\tfrac{1}{2 (4 r-1)}\right) \ln (r),
		\end{split}\\[.5cm]
		\begin{split}
			\sqrt{r}\,&\tilde{a}_4 = 80+\tfrac{4 (4 r-1)}{(r-1)^2}+ \tfrac{\pi ^2}{6} \left(\tfrac{4}{r}-6 r^2-7 r+27\right)
				 +\left(3 (4 r+3)+\tfrac{6 (7 r+1)}{(r-1)^2}-\tfrac{84}{r}\right) \tilde{I}(0,r) \\
			& -\left(\tfrac{4 (r+2)}{3 (r-1)^2}+2 (3 r+1)+\tfrac{26}{3 (r-4)}\right) \tilde{I}(r,1)
				 +\left(\tfrac{231-976 r}{4 (4 r-1)^2}+\tfrac{80}{r}+\tfrac{4}{(r-1)^2}-\tfrac{15}{4}\right) \tilde{I}(r,r) \\
			& +\left(\tfrac{57 r^2-136 r+67}{(r-1)^3}+\tfrac{13}{2 (4 r-1)}-\tfrac{5}{2}\right) \ln (r)
			 +\left(\tfrac{231-976 r}{8 (4 r-1)^2}+\tfrac{83 r+61}{6 (r-1)^2}-\tfrac{52}{3 (r-4)}-\tfrac{43}{8}\right) \ln ^2(r),
		\end{split}\\[.5cm]
		\begin{split}
			&\tilde{a}_5 = \tfrac{\pi ^2}{6} \left(\tfrac{4}{r}-6 r-1\right)-\tfrac{4}{r-1} +\left(\tfrac{2 (3 r-1)}{(r-1)^2}+\tfrac{2}{4 r-1}\right) \ln (r)
			+\left(\tfrac{5-22 r}{(4 r-1)^2}-\tfrac{28}{3 (r-4)}+\tfrac{16}{3 (r-1)}\right) \ln ^2(r)\\
			&+\left(\tfrac{18}{r-1}+12-\tfrac{12}{r}\right) \tilde{I}(0,r)
			 -\left(\tfrac{r+2}{3 (r-1)^2}+\tfrac{14}{3 (r-4)}+6\right) \tilde{I}(r,1)
			 +\left(\tfrac{8}{r}-\tfrac{2 (22 r-5)}{(4 r-1)^2}+\tfrac{1}{(r-1)^2}\right) \tilde{I}(r,r).
		\end{split}
	\end{align}
	\end{subequations}
The mass dependence of the coefficients for $C_{e\gamma}$ in Eq.~\eqref{eq:Ceg-2loop} is obtained from $a_i^{e\gamma} = a_i^{eB} - a_i^{eW}$.

\section{Analytic expressions for master integrals}\label{app:loop-functions}
For convenience we collect here the analytic expressions for the vacuum integrals appearing during the matching calculations.
The one-loop tadpole integral is given by the well-known expression
\begin{align}
	\begin{split}
		A_0(m^2) &\equiv \int \frac{d^Dk}{(2\pi)^D} \frac{-i(4\pi)^2\mu^{2\epsilon}}{k^2-m^2} = - m^2 \bigg(\frac{\bar\mu^2}{m^2}\bigg)^\epsilon \Gamma(\epsilon-1)e^{\gamma_E\epsilon} \\
		&= m^2 \Big[\tfrac{1}{\epsilon} + 1 - L(m^2) + \tfrac{\epsilon}{2} \Big(\tfrac{\pi^2}{6} + 1 + \big[1-L(m^2)\big]^2\Big)  + \O(\epsilon^2) \Big],
	\end{split}
\end{align}
where $L(m^n)\equiv n\,\ln(m/\bar\mu)$ and $\bar\mu^2 = 4\pi e^{-\gamma_E}\mu^2$. At two-loop order the two relevant topologies are
\begin{subequations}
	\begin{align}
		T_{14}(m_1^2,m_2^2) &\equiv \iint\frac{d^Dk_1}{(2\pi)^D}\frac{d^Dk_2}{(2\pi)^D} \frac{(4\pi)^4 \mu^{4\epsilon}}{[k_1^2-m_1^2][k_2^2-m_2^2]}, \\
		T_{134}(m_1^2,m_2^2,m_3^2) &\equiv \iint\frac{d^Dk_1}{(2\pi)^D}\frac{d^Dk_2}{(2\pi)^D} \frac{(4\pi)^4 \mu^{4\epsilon}}{[k_1^2-m_1^2][k_2^2-m_2^2][(k_1-k_2)^2-m_3^2]}.
	\end{align}
\end{subequations}
The first integral is just the product of two one-loop tadpoles and is given by
\begin{align}
	T_{14}(m_1^2,m_2^2) = - \frac{m_1^2m_2^2}{\epsilon^2} + \frac{2m_1^2 m_2^2}{\epsilon}  \Big[L(m_1)+L(m_2)-1\Big] + T_{14}^\text{fin}(m_1^2,m_2^2) + \O(\epsilon),
\end{align}
where
\begin{align}
	T_{14}^\text{fin}(m_1^2,m_2^2) = - m_1^2m_2^2 \Big(1+\tfrac{\pi^2}{6} + 2 \big[L(m_1)+L(m_2)-1\big]^2 \Big).
\end{align}
The second (sunset-vacuum) integral has been worked out e.g. in Refs.~\cite{Ford:1992pn,Davydychev:1992mt,Davydychev:1993pg,Davydychev:1995mq} and is given by
\begin{align}
	T_{134}(m_1^2,m_2^2,m_3^2) &= -\frac{1}{2\epsilon^2} \sum m_i^2 - \frac1\epsilon \sum_i m_i^2\Big[\tfrac{3}{2} - L(m_i^2)\Big] + T_{134}^\text{fin}(m_1^2,m_2^2,m_3^2) + \O(\epsilon).
\end{align}
The finite part can be written in closed form as
\begin{align}
	\begin{split}
		T_{134}^\text{fin}(m_1^2,m_2^2,m_3^2) = -\sum_i m_i^2 \bigg[ \frac{42+\pi^2}{12} + \Big(L(m_i^2)-3\Big)L(m_i^2)\bigg] - m_3^2 \, \tilde{I}(r_1,r_2),
	\end{split}
\end{align}
where the function $\tilde{I}(r_1,r_2)$ is given by
\begin{align}\label{eq:Itil}
	\begin{split}
		&\tilde{I}(r_1,r_2) = r_1\,\text{Li}_2\Big(1-\tfrac{r_2}{r_1}\Big) + r_2\,\text{Li}_2\Big(1-\tfrac{r_1}{r_2}\Big) -\frac{\pi^2}{6}\Delta \\
		&\qquad+ \sum_{i=1,2}\bigg[(2x_i-1)\mathrm{Li}_2(x_i)-\Delta\mathrm{Li}_2\Big(\frac{\Delta}{x_i}\Big)-\frac{\Delta}{2}\ln^2x_i-\ln(r_2)(1-x_i)\ln(1-x_i)\big],
	\end{split}
\end{align}
with $r_i = m_i^2/m_3^2$, $\Delta = r_2-r_1$ and $x_{1,2} = \frac{1}{2}\big(1-r_1+r_2\pm\sqrt{(1-r_1-r_2)^2 - 4r_1r_2}\big)$. 
Note that the results in App.~\ref{app:complete-result} were derived assuming $r\leq4$. For $r\geq 4$
the solution Eq.~\eqref{eq:Itil} is no longer valid expressed in the fact that $\tilde{I}(r,1)$ develops an imaginary part, while the actual
vacuum integral is real for any mass configuration. 
Fortunately, the real part of Eq.~\eqref{eq:Itil} remains correct and therefore the result for $r>4$ can be recovered by setting $\tilde{I}(r,1)\to\Re\tilde{I}(r,1)$.

\printbibliography

@article{Buchmuller:1985jz,
	author = "Buchmuller, W. and Wyler, D.",
	title = "{Effective Lagrangian Analysis of New Interactions and Flavor Conservation}",
	doi = "10.1016/0550-3213(86)90262-2",
	journal = "Nucl. Phys. B",
	volume = "268",
	pages = "621--653",
	year = "1986"
}

@article{Grzadkowski:2010es,
	author = "Grzadkowski, B. and Iskrzynski, M. and Misiak, M. and Rosiek, J.",
	title = "{Dimension-Six Terms in the Standard Model Lagrangian}",
	eprint = "1008.4884",
	archivePrefix = "arXiv",
	primaryClass = "hep-ph",
	doi = "10.1007/JHEP10(2010)085",
	journal = "JHEP",
	volume = "10",
	pages = "085",
	year = "2010"
}

@article{deBlas:2017xtg,
	author = "de Blas, J. and Criado, J. C. and Perez-Victoria, M. and Santiago, J.",
	title = "{Effective description of general extensions of the Standard Model: the complete tree-level dictionary}",
	eprint = "1711.10391",
	archivePrefix = "arXiv",
	primaryClass = "hep-ph",
	doi = "10.1007/JHEP03(2018)109",
	journal = "JHEP",
	volume = "03",
	pages = "109",
	year = "2018"
}

@article{Czarnecki:1995sz,
	author = "Czarnecki, Andrzej and Krause, Bernd and Marciano, William J.",
	title = "{Electroweak corrections to the muon anomalous magnetic moment}",
	eprint = "hep-ph/9512369",
	archivePrefix = "arXiv",
	doi = "10.1103/PhysRevLett.76.3267",
	journal = "Phys. Rev. Lett.",
	volume = "76",
	pages = "3267--3270",
	year = "1996"
}

@article{Carmona:2021xtq,
	author = "Carmona, Adrian and Lazopoulos, Achilleas and Olgoso, Pablo and Santiago, Jose",
	title = "{Matchmakereft: automated tree-level and one-loop matching}",
	eprint = "2112.10787",
	archivePrefix = "arXiv",
	primaryClass = "hep-ph",
	doi = "10.21468/SciPostPhys.12.6.198",
	journal = "SciPost Phys.",
	volume = "12",
	number = "6",
	pages = "198",
	year = "2022"
}

@article{Guedes:2023azv,
	author = "Guedes, Guilherme and Olgoso, Pablo and Santiago, Jos{\'e}",
	title = "{Towards the one loop IR/UV dictionary in the SMEFT: One loop generated operators from new scalars and fermions}",
	eprint = "2303.16965",
	archivePrefix = "arXiv",
	primaryClass = "hep-ph",
	doi = "10.21468/SciPostPhys.15.4.143",
	journal = "SciPost Phys.",
	volume = "15",
	number = "4",
	pages = "143",
	year = "2023"
}

@article{Fuentes-Martin:2023ljp,
	author = {Fuentes-Mart{\'\i}n, Javier and Palavri{\'c}, Ajdin and Thomsen, Anders Eller},
	title = "{Functional matching and renormalization group equations at two-loop order}",
	eprint = "2311.13630",
	archivePrefix = "arXiv",
	primaryClass = "hep-ph",
	doi = "10.1016/j.physletb.2024.138557",
	journal = "Phys. Lett. B",
	volume = "851",
	pages = "138557",
	year = "2024"
}

@article{Panico:2018hal,
	author = "Panico, Giuliano and Pomarol, Alex and Riembau, Marc",
	title = "{EFT approach to the electron Electric Dipole Moment at the two-loop level}",
	eprint = "1810.09413",
	archivePrefix = "arXiv",
	primaryClass = "hep-ph",
	doi = "10.1007/JHEP04(2019)090",
	journal = "JHEP",
	volume = "04",
	pages = "090",
	year = "2019"
}

@article{Alonso:2013hga,
	author = "Alonso, Rodrigo and Jenkins, Elizabeth E. and Manohar, Aneesh V. and Trott, Michael",
	title = "{Renormalization Group Evolution of the Standard Model Dimension Six Operators III: Gauge Coupling Dependence and Phenomenology}",
	eprint = "1312.2014",
	archivePrefix = "arXiv",
	primaryClass = "hep-ph",
	reportNumber = "CERN-PH-TH-2013-305, CERN-PH-TH/2013-305",
	doi = "10.1007/JHEP04(2014)159",
	journal = "JHEP",
	volume = "04",
	pages = "159",
	year = "2014"
}

@article{Jenkins:2013zja,
	author = "Jenkins, Elizabeth E. and Manohar, Aneesh V. and Trott, Michael",
	title = "{Renormalization Group Evolution of the Standard Model Dimension Six Operators I: Formalism and lambda Dependence}",
	eprint = "1308.2627",
	archivePrefix = "arXiv",
	primaryClass = "hep-ph",
	doi = "10.1007/JHEP10(2013)087",
	journal = "JHEP",
	volume = "10",
	pages = "087",
	year = "2013"
}

@article{Jenkins:2013wua,
	author = "Jenkins, Elizabeth E. and Manohar, Aneesh V. and Trott, Michael",
	title = "{Renormalization Group Evolution of the Standard Model Dimension Six Operators II: Yukawa Dependence}",
	eprint = "1310.4838",
	archivePrefix = "arXiv",
	primaryClass = "hep-ph",
	reportNumber = "CERN-PH-TH/2015-247",
	doi = "10.1007/JHEP01(2014)035",
	journal = "JHEP",
	volume = "01",
	pages = "035",
	year = "2014"
}

@article{Jenkins:2017dyc,
	author = "Jenkins, Elizabeth E. and Manohar, Aneesh V. and Stoffer, Peter",
	title = "{Low-Energy Effective Field Theory below the Electroweak Scale: Anomalous Dimensions}",
	eprint = "1711.05270",
	archivePrefix = "arXiv",
	primaryClass = "hep-ph",
	doi = "10.1007/JHEP01(2018)084",
	journal = "JHEP",
	volume = "01",
	pages = "084",
	year = "2018",
	note = "[Erratum: JHEP 12, 042 (2023)]"
}

@article{Degrassi:1998es,
	author = "Degrassi, G. and Giudice, G. F.",
	title = "{QED logarithms in the electroweak corrections to the muon anomalous magnetic moment}",
	eprint = "hep-ph/9803384",
	archivePrefix = "arXiv",
	reportNumber = "CERN-TH-98-086, CERN-TH-98-86",
	doi = "10.1103/PhysRevD.58.053007",
	journal = "Phys. Rev. D",
	volume = "58",
	pages = "053007",
	year = "1998"
}

@article{delAguila:1982fs,
	author = "del Aguila, F. and Bowick, Mark J.",
	title = "{The Possibility of New Fermions With $\Delta$ I = 0 Mass}",
	reportNumber = "UFTP-82-19",
	doi = "10.1016/0550-3213(83)90316-4",
	journal = "Nucl. Phys. B",
	volume = "224",
	pages = "107",
	year = "1983"
}

@article{Langacker:1988ur,
	author = "Langacker, Paul and London, David",
	title = "{Mixing Between Ordinary and Exotic Fermions}",
	reportNumber = "DESY-88-043",
	doi = "10.1103/PhysRevD.38.886",
	journal = "Phys. Rev. D",
	volume = "38",
	pages = "886",
	year = "1988"
}

@article{Frampton:1999xi,
	author = "Frampton, Paul H. and Hung, P. Q. and Sher, Marc",
	title = "{Quarks and leptons beyond the third generation}",
	eprint = "hep-ph/9903387",
	archivePrefix = "arXiv",
	reportNumber = "IFP-759-UNCA, WM-99-104",
	doi = "10.1016/S0370-1573(99)00095-2",
	journal = "Phys. Rept.",
	volume = "330",
	pages = "263",
	year = "2000"
}

@article{Gursey:1975ki,
	author = "Gursey, F. and Ramond, Pierre and Sikivie, P.",
	title = "{A Universal Gauge Theory Model Based on E6}",
	reportNumber = "YALE-3075-118",
	doi = "10.1016/0370-2693(76)90417-2",
	journal = "Phys. Lett. B",
	volume = "60",
	pages = "177--180",
	year = "1976"
}

@article{Hewett:1988xc,
	author = "Hewett, JoAnne L. and Rizzo, Thomas G.",
	title = "{Low-Energy Phenomenology of Superstring Inspired E(6) Models}",
	reportNumber = "MAD-PH-446, IS-J-3005",
	doi = "10.1016/0370-1573(89)90071-9",
	journal = "Phys. Rept.",
	volume = "183",
	pages = "193",
	year = "1989"
}

@article{Dermisek:2012ke,
	author = "Dermisek, Radovan",
	title = "{Unification of gauge couplings in the standard model with extra vectorlike families}",
	eprint = "1212.3035",
	archivePrefix = "arXiv",
	primaryClass = "hep-ph",
	doi = "10.1103/PhysRevD.87.055008",
	journal = "Phys. Rev. D",
	volume = "87",
	number = "5",
	pages = "055008",
	year = "2013"
}

@article{Dermisek:2012as,
	author = "Dermisek, Radovan",
	title = "{Insensitive Unification of Gauge Couplings}",
	eprint = "1204.6533",
	archivePrefix = "arXiv",
	primaryClass = "hep-ph",
	doi = "10.1016/j.physletb.2012.06.037",
	journal = "Phys. Lett. B",
	volume = "713",
	pages = "469--472",
	year = "2012"
}

@article{Dorsner:2014wva,
	author = "Dorsner, Ilja and Fajfer, Svjetlana and Mustac, Ivana",
	title = "{Light vector-like fermions in a minimal SU(5) setup}",
	eprint = "1401.6870",
	archivePrefix = "arXiv",
	primaryClass = "hep-ph",
	doi = "10.1103/PhysRevD.89.115004",
	journal = "Phys. Rev. D",
	volume = "89",
	number = "11",
	pages = "115004",
	year = "2014"
}

@article{Bhattacherjee:2017cxh,
	author = "Bhattacherjee, Biplob and Byakti, Pritibhajan and Kushwaha, Ashwani and Vempati, Sudhir K",
	title = "{Unification with Vector-like fermions and signals at LHC}",
	eprint = "1702.06417",
	archivePrefix = "arXiv",
	primaryClass = "hep-ph",
	doi = "10.1007/JHEP05(2018)090",
	journal = "JHEP",
	volume = "05",
	pages = "090",
	year = "2018"
}

@article{Kowalska:2019qxm,
	author = "Kowalska, Kamila and Kumar, Dinesh",
	title = "{Road map through the desert: unification with vector-like fermions}",
	eprint = "1910.00847",
	archivePrefix = "arXiv",
	primaryClass = "hep-ph",
	doi = "10.1007/JHEP12(2019)094",
	journal = "JHEP",
	volume = "12",
	pages = "094",
	year = "2019"
}

@article{Gherghetta:2000qt,
	author = "Gherghetta, Tony and Pomarol, Alex",
	title = "{Bulk fields and supersymmetry in a slice of AdS}",
	eprint = "hep-ph/0003129",
	archivePrefix = "arXiv",
	reportNumber = "CERN-TH-2000-081, UNIL-IPT-00-06",
	doi = "10.1016/S0550-3213(00)00392-8",
	journal = "Nucl. Phys. B",
	volume = "586",
	pages = "141--162",
	year = "2000"
}

@article{Agashe:2008fe,
	author = "Agashe, Kaustubh and Okui, Takemichi and Sundrum, Raman",
	title = "{A Common Origin for Neutrino Anarchy and Charged Hierarchies}",
	eprint = "0810.1277",
	archivePrefix = "arXiv",
	primaryClass = "hep-ph",
	reportNumber = "UMD-PP-08-019",
	doi = "10.1103/PhysRevLett.102.101801",
	journal = "Phys. Rev. Lett.",
	volume = "102",
	pages = "101801",
	year = "2009"
}

@article{delAguila:2008pw,
	author = "del Aguila, F. and de Blas, J. and Perez-Victoria, M.",
	title = "{Effects of new leptons in Electroweak Precision Data}",
	eprint = "0803.4008",
	archivePrefix = "arXiv",
	primaryClass = "hep-ph",
	reportNumber = "UG-FT-224-08, CAFPE-94-08",
	doi = "10.1103/PhysRevD.78.013010",
	journal = "Phys. Rev. D",
	volume = "78",
	pages = "013010",
	year = "2008"
}

@article{Baspehlivan:2022qet,
	author = "Baspehlivan, Feyza and Dagli, Burak and Delialioglu, Osman Emre and Sultansoy, Saleh",
	title = "{Why should we search for vector-like leptons?}",
	eprint = "2201.08251",
	archivePrefix = "arXiv",
	primaryClass = "hep-ph",
	month = "1",
	year = "2022"
}

@article{Huber:2000ie,
	author = "Huber, Stephan J. and Shafi, Qaisar",
	title = "{Fermion masses, mixings and proton decay in a Randall-Sundrum model}",
	eprint = "hep-ph/0010195",
	archivePrefix = "arXiv",
	reportNumber = "BA-00-42",
	doi = "10.1016/S0370-2693(00)01399-X",
	journal = "Phys. Lett. B",
	volume = "498",
	pages = "256--262",
	year = "2001"
}

@article{Ishiwata:2011hr,
	author = "Ishiwata, Koji and Wise, Mark B.",
	title = "{Higgs Properties and Fourth Generation Leptons}",
	eprint = "1107.1490",
	archivePrefix = "arXiv",
	primaryClass = "hep-ph",
	reportNumber = "CALT-68-2844",
	doi = "10.1103/PhysRevD.84.055025",
	journal = "Phys. Rev. D",
	volume = "84",
	pages = "055025",
	year = "2011"
}

@article{Ishiwata:2013gma,
	author = "Ishiwata, Koji and Wise, Mark B.",
	title = "{Phenomenology of heavy vectorlike leptons}",
	eprint = "1307.1112",
	archivePrefix = "arXiv",
	primaryClass = "hep-ph",
	reportNumber = "CALT-68-2943",
	doi = "10.1103/PhysRevD.88.055009",
	journal = "Phys. Rev. D",
	volume = "88",
	number = "5",
	pages = "055009",
	year = "2013"
}

@article{Ishiwata:2015cga,
	author = "Ishiwata, Koji and Ligeti, Zoltan and Wise, Mark B.",
	title = "{New Vector-Like Fermions and Flavor Physics}",
	eprint = "1506.03484",
	archivePrefix = "arXiv",
	primaryClass = "hep-ph",
	reportNumber = "CALT-TH-2015-029, KANAZAWA-15-07",
	doi = "10.1007/JHEP10(2015)027",
	journal = "JHEP",
	volume = "10",
	pages = "027",
	year = "2015"
}

@article{Goudelis:2011un,
	author = "Goudelis, Andreas and Lebedev, Oleg and Park, Jae-hyeon",
	title = "{Higgs-induced lepton flavor violation}",
	eprint = "1111.1715",
	archivePrefix = "arXiv",
	primaryClass = "hep-ph",
	reportNumber = "DESY-11-194",
	doi = "10.1016/j.physletb.2011.12.059",
	journal = "Phys. Lett. B",
	volume = "707",
	pages = "369--374",
	year = "2012"
}

@article{Kannike:2011ng,
	author = "Kannike, Kristjan and Raidal, Martti and Straub, David M. and Strumia, Alessandro",
	title = "{Anthropic solution to the magnetic muon anomaly: the charged see-saw}",
	eprint = "1111.2551",
	archivePrefix = "arXiv",
	primaryClass = "hep-ph",
	doi = "10.1007/JHEP02(2012)106",
	journal = "JHEP",
	volume = "02",
	pages = "106",
	year = "2012",
	note = "[Erratum: JHEP 10, 136 (2012)]"
}

@article{Kearney:2012zi,
	author = "Kearney, John and Pierce, Aaron and Weiner, Neal",
	title = "{Vectorlike Fermions and Higgs Couplings}",
	eprint = "1207.7062",
	archivePrefix = "arXiv",
	primaryClass = "hep-ph",
	reportNumber = "MCTP-12-17",
	doi = "10.1103/PhysRevD.86.113005",
	journal = "Phys. Rev. D",
	volume = "86",
	pages = "113005",
	year = "2012"
}

@article{Dermisek:2013gta,
	author = "Dermisek, Radovan and Raval, Aditi",
	title = "{Explanation of the Muon g-2 Anomaly with Vectorlike Leptons and its Implications for Higgs Decays}",
	eprint = "1305.3522",
	archivePrefix = "arXiv",
	primaryClass = "hep-ph",
	doi = "10.1103/PhysRevD.88.013017",
	journal = "Phys. Rev. D",
	volume = "88",
	pages = "013017",
	year = "2013"
}

@article{Falkowski:2013jya,
	author = "Falkowski, Adam and Straub, David M. and Vicente, Avelino",
	title = "{Vector-like leptons: Higgs decays and collider phenomenology}",
	eprint = "1312.5329",
	archivePrefix = "arXiv",
	primaryClass = "hep-ph",
	reportNumber = "LPT-ORSAY-13-141",
	doi = "10.1007/JHEP05(2014)092",
	journal = "JHEP",
	volume = "05",
	pages = "092",
	year = "2014"
}

@article{Daberstiel:2026wxe,
	author = {Daberstiel, Gregor and M{\"o}hling, Kilian and St{\"o}ckinger, Dominik and St{\"o}ckinger-Kim, Hyejung},
	title = "{From Higgs physics to lepton flavour violation: current bounds and future prospects for vector-like lepton models}",
	eprint = "2603.21414",
	archivePrefix = "arXiv",
	primaryClass = "hep-ph",
	doi = "10.1007/JHEP07(2026)259",
	journal = "JHEP",
	volume = "07",
	pages = "259",
	year = "2026"
}

@article{Ellis:2014dza,
	author = "Ellis, Sebastian A. R. and Godbole, Rohini M. and Gopalakrishna, Shrihari and Wells, James D.",
	title = "{Survey of vector-like fermion extensions of the Standard Model and their phenomenological implications}",
	eprint = "1404.4398",
	archivePrefix = "arXiv",
	primaryClass = "hep-ph",
	doi = "10.1007/JHEP09(2014)130",
	journal = "JHEP",
	volume = "09",
	pages = "130",
	year = "2014"
}

@article{Poh:2017tfo,
	author = "Poh, Zijie and Raby, Stuart",
	title = "{Vectorlike leptons: Muon g-2 anomaly, lepton flavor violation, Higgs boson decays, and lepton nonuniversality}",
	eprint = "1705.07007",
	archivePrefix = "arXiv",
	primaryClass = "hep-ph",
	doi = "10.1103/PhysRevD.96.015032",
	journal = "Phys. Rev. D",
	volume = "96",
	number = "1",
	pages = "015032",
	year = "2017"
}

@article{Raby:2017igl,
	author = "Raby, Stuart and Trautner, Andreas",
	title = "{Vectorlike chiral fourth family to explain muon anomalies}",
	eprint = "1712.09360",
	archivePrefix = "arXiv",
	primaryClass = "hep-ph",
	doi = "10.1103/PhysRevD.97.095006",
	journal = "Phys. Rev. D",
	volume = "97",
	number = "9",
	pages = "095006",
	year = "2018"
}

@article{Chala:2020odv,
	author = "Chala, Mikael and Koz{\'o}w, Pawe{\l} and Ramos, Maria and Titov, Arsenii",
	title = "{Effective field theory for vector-like leptons and its collider signals}",
	eprint = "2005.09655",
	archivePrefix = "arXiv",
	primaryClass = "hep-ph",
	doi = "10.1016/j.physletb.2020.135752",
	journal = "Phys. Lett. B",
	volume = "809",
	pages = "135752",
	year = "2020"
}

@article{Crivellin:2020ebi,
	author = "Crivellin, Andreas and Kirk, Fiona and Manzari, Claudio Andrea and Montull, Marc",
	title = "{Global Electroweak Fit and Vector-Like Leptons in Light of the Cabibbo Angle Anomaly}",
	eprint = "2008.01113",
	archivePrefix = "arXiv",
	primaryClass = "hep-ph",
	reportNumber = "CERN-TH-2020-128, PSI-PR-20-11, UZ-TH 26/20",
	doi = "10.1007/JHEP12(2020)166",
	journal = "JHEP",
	volume = "12",
	pages = "166",
	year = "2020"
}

@article{Arkani-Hamed:2021xlp,
	author = "Arkani-Hamed, Nima and Harigaya, Keisuke",
	title = "{Naturalness and the muon magnetic moment}",
	eprint = "2106.01373",
	archivePrefix = "arXiv",
	primaryClass = "hep-ph",
	doi = "10.1007/JHEP09(2021)025",
	journal = "JHEP",
	volume = "09",
	pages = "025",
	year = "2021"
}

@article{Hamaguchi:2022byw,
	author = "Hamaguchi, Koichi and Nagata, Natsumi and Osaki, Genta and Tseng, Shih-Yen",
	title = "{Probing new physics in the vector-like lepton model by lepton electric dipole moments}",
	eprint = "2211.16800",
	archivePrefix = "arXiv",
	primaryClass = "hep-ph",
	doi = "10.1007/JHEP01(2023)100",
	journal = "JHEP",
	volume = "01",
	pages = "100",
	year = "2023"
}

@article{Erdelyi:2025axy,
	author = {Erdelyi, Barbara Anna and Gr{\"o}ber, Ramona and Selimovic, Nudzeim},
	title = "{Probing new physics with the electron Yukawa coupling}",
	eprint = "2501.07628",
	archivePrefix = "arXiv",
	primaryClass = "hep-ph",
	doi = "10.1007/JHEP05(2025)135",
	journal = "JHEP",
	volume = "05",
	pages = "135",
	year = "2025"
}

@article{Dermisek:2022aec,
	author = "Dermisek, Radovan and Hermanek, Keith and McGinnis, Navin and Yoon, Sangsik",
	title = "{Ellipse of Muon Dipole Moments}",
	eprint = "2205.14243",
	archivePrefix = "arXiv",
	primaryClass = "hep-ph",
	doi = "10.1103/PhysRevLett.129.221801",
	journal = "Phys. Rev. Lett.",
	volume = "129",
	number = "22",
	pages = "221801",
	year = "2022"
}

@article{Dermisek:2023nhe,
	author = "Dermisek, Radovan and Hermanek, Keith and McGinnis, Navin and Yoon, Sangsik",
	title = "{Effective field theory of chirally enhanced muon mass and dipole operators}",
	eprint = "2302.14144",
	archivePrefix = "arXiv",
	primaryClass = "hep-ph",
	doi = "10.1103/PhysRevD.107.095043",
	journal = "Phys. Rev. D",
	volume = "107",
	number = "9",
	pages = "095043",
	year = "2023"
}

@article{Mohling:2024qvk,
	author = {M{\"o}hling, Kilian and St{\"o}ckinger, Dominik and St{\"o}ckinger-Kim, Hyejung},
	title = "{On-shell renormalization with vector-like leptons, one-loop muon{\textendash}Higgs coupling and muon g {\ensuremath{-}} 2}",
	eprint = "2407.09421",
	archivePrefix = "arXiv",
	primaryClass = "hep-ph",
	doi = "10.1007/JHEP10(2024)170",
	journal = "JHEP",
	volume = "10",
	pages = "170",
	year = "2024"
}

@article{Adhikary:2024esf,
	author = "Adhikary, Amit and Olechowski, Marek and Rosiek, Janusz and Ryczkowski, Michal",
	title = "{Theoretical constraints on models with vectorlike fermions}",
	eprint = "2406.16050",
	archivePrefix = "arXiv",
	primaryClass = "hep-ph",
	doi = "10.1103/PhysRevD.110.075029",
	journal = "Phys. Rev. D",
	volume = "110",
	number = "7",
	pages = "075029",
	year = "2024"
}

@article{Cingiloglu:2024vdh,
	author = "Cingiloglu, Kivanc Y. and Frank, Mariana",
	title = "{Stability of the standard model vacuum with vectorlike leptons: A critical examination}",
	eprint = "2408.10898",
	archivePrefix = "arXiv",
	primaryClass = "hep-ph",
	doi = "10.1103/PhysRevD.111.016025",
	journal = "Phys. Rev. D",
	volume = "111",
	number = "1",
	pages = "016025",
	year = "2025"
}

@article{Gopalakrishna:2018uxn,
	author = "Gopalakrishna, Shrihari and Velusamy, Arunprasath",
	title = "{Higgs vacuum stability with vectorlike fermions}",
	eprint = "1812.11303",
	archivePrefix = "arXiv",
	primaryClass = "hep-ph",
	doi = "10.1103/PhysRevD.99.115020",
	journal = "Phys. Rev. D",
	volume = "99",
	number = "11",
	pages = "115020",
	year = "2019"
}

@article{Raju:2022zlv,
	author = "Raju, Md. and Mukherjee, Abhi and Saha, Jyoti Prasad",
	title = "{Investigation of $(g-2)_{\mu }$ anomaly in the $\mu $-specific 2HDM with vector like leptons and the phenomenological implications}",
	eprint = "2207.02825",
	archivePrefix = "arXiv",
	primaryClass = "hep-ph",
	doi = "10.1140/epjc/s10052-023-11595-8",
	journal = "Eur. Phys. J. C",
	volume = "83",
	number = "5",
	pages = "429",
	year = "2023"
}

@article{Chun:2020uzw,
	author = "Chun, Eung Jin and Mondal, Tanmoy",
	title = "{Explaining $g-2$ anomalies in two Higgs doublet model with vector-like leptons}",
	eprint = "2009.08314",
	archivePrefix = "arXiv",
	primaryClass = "hep-ph",
	reportNumber = "KIAS - P20050",
	doi = "10.1007/JHEP11(2020)077",
	journal = "JHEP",
	volume = "11",
	pages = "077",
	year = "2020"
}

@article{Dermisek:2023tgq,
	author = "Dermisek, Radovan and Hermanek, Keith and McGinnis, Navin and Yoon, Sangsik",
	title = "{Predictions for muon electric and magnetic dipole moments from h{\textrightarrow}{\ensuremath{\mu}}+{\ensuremath{\mu}}- in two-Higgs-doublet models with new leptons}",
	eprint = "2306.13212",
	archivePrefix = "arXiv",
	primaryClass = "hep-ph",
	doi = "10.1103/PhysRevD.108.055019",
	journal = "Phys. Rev. D",
	volume = "108",
	number = "5",
	pages = "055019",
	year = "2023"
}

@article{Fuentes-Martin:2022jrf,
	author = {Fuentes-Mart{\'\i}n, Javier and K{\"o}nig, Matthias and Pag{\`e}s, Julie and Thomsen, Anders Eller and Wilsch, Felix},
	title = "{A proof of concept for matchete: an automated tool for matching effective theories}",
	eprint = "2212.04510",
	archivePrefix = "arXiv",
	primaryClass = "hep-ph",
	reportNumber = "MITP-22-105, TUM-HEP-1443/22, ZU-TH-58/22",
	doi = "10.1140/epjc/s10052-023-11726-1",
	journal = "Eur. Phys. J. C",
	volume = "83",
	number = "7",
	pages = "662",
	year = "2023"
}

@article{Shtabovenko:2023idz,
	author = "Shtabovenko, Vladyslav and Mertig, Rolf and Orellana, Frederik",
	title = "{FeynCalc 10: Do multiloop integrals dream of computer codes?}",
	eprint = "2312.14089",
	archivePrefix = "arXiv",
	primaryClass = "hep-ph",
	reportNumber = "P3H-23-089, TTP23-056, SI-HEP-2023-27",
	doi = "10.1016/j.cpc.2024.109357",
	journal = "Comput. Phys. Commun.",
	volume = "306",
	pages = "109357",
	year = "2025"
}

@article{Shtabovenko:2020gxv,
	author = "Shtabovenko, Vladyslav and Mertig, Rolf and Orellana, Frederik",
	title = "{FeynCalc 9.3: New features and improvements}",
	eprint = "2001.04407",
	archivePrefix = "arXiv",
	primaryClass = "hep-ph",
	reportNumber = "P3H-20-002, TTP19-020, TUM-EFT 130/19",
	doi = "10.1016/j.cpc.2020.107478",
	journal = "Comput. Phys. Commun.",
	volume = "256",
	pages = "107478",
	year = "2020"
}

@article{Shtabovenko:2016sxi,
	author = "Shtabovenko, Vladyslav and Mertig, Rolf and Orellana, Frederik",
	title = "{New Developments in FeynCalc 9.0}",
	eprint = "1601.01167",
	archivePrefix = "arXiv",
	primaryClass = "hep-ph",
	reportNumber = "TUM-EFT-71-15",
	doi = "10.1016/j.cpc.2016.06.008",
	journal = "Comput. Phys. Commun.",
	volume = "207",
	pages = "432--444",
	year = "2016"
}

@article{Mertig:1990an,
	author = "Mertig, R. and Bohm, M. and Denner, Ansgar",
	title = "{FEYN CALC: Computer algebraic calculation of Feynman amplitudes}",
	reportNumber = "PRINT-90-0639 (WURZBURG)",
	doi = "10.1016/0010-4655(91)90130-D",
	journal = "Comput. Phys. Commun.",
	volume = "64",
	pages = "345--359",
	year = "1991"
}

@article{Christensen:2008py,
	author = "Christensen, Neil D. and Duhr, Claude",
	title = "{FeynRules - Feynman rules made easy}",
	eprint = "0806.4194",
	archivePrefix = "arXiv",
	primaryClass = "hep-ph",
	reportNumber = "MSUHEP-080616, CP3-08-20",
	doi = "10.1016/j.cpc.2009.02.018",
	journal = "Comput. Phys. Commun.",
	volume = "180",
	pages = "1614--1641",
	year = "2009"
}

@article{Alloul:2013bka,
	author = "Alloul, Adam and Christensen, Neil D. and Degrande, C{\'e}line and Duhr, Claude and Fuks, Benjamin",
	title = "{FeynRules  2.0 - A complete toolbox for tree-level phenomenology}",
	eprint = "1310.1921",
	archivePrefix = "arXiv",
	primaryClass = "hep-ph",
	reportNumber = "CERN-PH-TH-2013-239, MCNET-13-14, IPPP-13-71, DCPT-13-142, PITT-PACC-1308",
	doi = "10.1016/j.cpc.2014.04.012",
	journal = "Comput. Phys. Commun.",
	volume = "185",
	pages = "2250--2300",
	year = "2014"
}

@article{Hahn:2000kx,
	author = "Hahn, Thomas",
	title = "{Generating Feynman diagrams and amplitudes with FeynArts 3}",
	eprint = "hep-ph/0012260",
	archivePrefix = "arXiv",
	reportNumber = "KA-TP-23-2000",
	doi = "10.1016/S0010-4655(01)00290-9",
	journal = "Comput. Phys. Commun.",
	volume = "140",
	pages = "418--431",
	year = "2001"
}

@article{Shtabovenko:2016whf,
	author = "Shtabovenko, Vladyslav",
	title = "{FeynHelpers: Connecting FeynCalc to FIRE and Package-X}",
	eprint = "1611.06793",
	archivePrefix = "arXiv",
	primaryClass = "physics.comp-ph",
	reportNumber = "TUM-EFT-75-15",
	doi = "10.1016/j.cpc.2017.04.014",
	journal = "Comput. Phys. Commun.",
	volume = "218",
	pages = "48--65",
	year = "2017"
}

@article{Shtabovenko:2025lxq,
	author = "Shtabovenko, Vladyslav",
	title = "{FeynCalc 10.2 and FeynHelpers 2: Multiloop calculations streamlined}",
	eprint = "2512.19858",
	archivePrefix = "arXiv",
	primaryClass = "hep-ph",
	reportNumber = "P3H-25-112, SI-HEP-2025-31",
	month = "12",
	year = "2025"
}

@article{Smirnov:2019qkx,
	author = "Smirnov, A. V. and Chukharev, F. S.",
	title = "{FIRE6: Feynman Integral REduction with modular arithmetic}",
	eprint = "1901.07808",
	archivePrefix = "arXiv",
	primaryClass = "hep-ph",
	reportNumber = "TTP19-006",
	doi = "10.1016/j.cpc.2019.106877",
	journal = "Comput. Phys. Commun.",
	volume = "247 ",
	pages = "106877",
	year = "2020"
}

@article{Smirnov:2025prc,
	author = "Smirnov, Alexander V. and Zeng, Mao",
	title = "{FIRE 7: Automatic Reduction with Modular Approach}",
	eprint = "2510.07150",
	archivePrefix = "arXiv",
	primaryClass = "hep-ph",
	month = "10",
	year = "2025"
}

@article{Davydychev:1992mt,
	author = "Davydychev, Andrei I. and Tausk, J. B.",
	title = "{Two loop selfenergy diagrams with different masses and the momentum expansion}",
	reportNumber = "INLO-PUB-11-92",
	doi = "10.1016/0550-3213(93)90338-P",
	journal = "Nucl. Phys. B",
	volume = "397",
	pages = "123--142",
	year = "1993"
}

@article{Davydychev:1993pg,
	author = "Davydychev, Andrei I. and Smirnov, Vladimir A. and Tausk, J. B.",
	title = "{Large momentum expansion of two loop selfenergy diagrams with arbitrary masses}",
	eprint = "hep-ph/9307371",
	archivePrefix = "arXiv",
	reportNumber = "INLO-PUB-5-93",
	doi = "10.1016/0550-3213(93)90436-S",
	journal = "Nucl. Phys. B",
	volume = "410",
	pages = "325--342",
	year = "1993"
}

@article{Davydychev:1995mq,
	author = "Davydychev, Andrei I. and Tausk, J. B.",
	title = "{A Magic connection between massive and massless diagrams}",
	eprint = "hep-ph/9504431",
	archivePrefix = "arXiv",
	reportNumber = "MZ-TH-95-14, BERGEN-1995-07",
	doi = "10.1103/PhysRevD.53.7381",
	journal = "Phys. Rev. D",
	volume = "53",
	pages = "7381--7384",
	year = "1996"
}

@article{Ford:1992pn,
	author = "Ford, C. and Jack, I. and Jones, D. R. T.",
	title = "{The Standard model effective potential at two loops}",
	eprint = "hep-ph/0111190",
	archivePrefix = "arXiv",
	reportNumber = "NSF-ITP-92-21, LTH-281",
	doi = "10.1016/0550-3213(92)90165-8",
	journal = "Nucl. Phys. B",
	volume = "387",
	pages = "373--390",
	year = "1992",
	note = "[Erratum: Nucl.Phys.B 504, 551--552 (1997)]"
}

@article{Muong-2:2025xyk,
	author = "Aguillard, D. P. and others",
	collaboration = "Muon g-2",
	title = "{Measurement of the Positive Muon Anomalous Magnetic Moment to 127 ppb}",
	eprint = "2506.03069",
	archivePrefix = "arXiv",
	primaryClass = "hep-ex",
	doi = "10.1103/7clf-sm2v",
	journal = "Phys. Rev. Lett.",
	volume = "135",
	number = "10",
	pages = "101802",
	year = "2025"
}

@article{Aliberti:2025beg,
	author = "Aliberti, R. and others",
	title = "{The anomalous magnetic moment of the muon in the Standard Model: an update}",
	eprint = "2505.21476",
	archivePrefix = "arXiv",
	primaryClass = "hep-ph",
	doi = "10.1016/j.physrep.2025.08.002",
	journal = "Phys. Rept.",
	volume = "1143",
	pages = "1--158",
	year = "2025"
}

@article{Aoyama:2020ynm,
	author = "Aoyama, T. and others",
	title = "{The anomalous magnetic moment of the muon in the Standard Model}",
	eprint = "2006.04822",
	archivePrefix = "arXiv",
	primaryClass = "hep-ph",
	reportNumber = "FERMILAB-PUB-20-207-T, INT-PUB-20-021, KEK Preprint 2020-5,
	MITP/20-028, KEK Preprint 2020-5, MITP/20-028, CERN-TH-2020-075, IFT-UAM/CSIC-20-74, LMU-ASC 18/20, LTH 1234,
	LU TP 20-20, LTH 1234, LU TP 20-20, MAN/HEP/2020/003, PSI-PR-20-06, UWThPh 2020-14, ZU-TH 18/20",
	doi = "10.1016/j.physrep.2020.07.006",
	journal = "Phys. Rept.",
	volume = "887",
	pages = "1--166",
	year = "2020"
}

@article{Roussy:2022cmp,
	author = "Roussy, Tanya S. and others",
	title = "{An improved bound on the electron's electric dipole moment}",
	eprint = "2212.11841",
	archivePrefix = "arXiv",
	primaryClass = "physics.atom-ph",
	doi = "10.1126/science.adg4084",
	journal = "Science",
	volume = "381",
	number = "6653",
	pages = "46",
	year = "2023"
}

@article{CMS:2020xwi,
	author = "Sirunyan, Albert M and others",
	collaboration = "CMS",
	title = "{Evidence for Higgs boson decay to a pair of muons}",
	eprint = "2009.04363",
	archivePrefix = "arXiv",
	primaryClass = "hep-ex",
	doi = "10.1007/JHEP01(2021)148",
	journal = "JHEP",
	volume = "01",
	pages = "148",
	year = "2021"
}

@article{ATLAS:2025coj,
	author = "Aad, Georges and others",
	collaboration = "ATLAS",
	title = "{Evidence for the dimuon decay of the Higgs boson in pp collisions with the ATLAS detector}",
	eprint = "2507.03595",
	archivePrefix = "arXiv",
	primaryClass = "hep-ex",
	doi = "10.1103/gzdh-p159",
	journal = "Phys. Rev. Lett.",
	volume = "135",
	number = "23",
	pages = "231802",
	year = "2025"
}

@article{ATLAS:2022vkf,
	author = "Aad, Georges and others",
	collaboration = "ATLAS",
	title = "{A detailed map of Higgs boson interactions by the ATLAS experiment ten years after the discovery}",
	eprint = "2207.00092",
	archivePrefix = "arXiv",
	primaryClass = "hep-ex",
	doi = "10.1038/s41586-022-04893-w",
	journal = "Nature",
	volume = "607",
	number = "7917",
	pages = "52--59",
	year = "2022"
}

@article{CMS:2022dwd,
	author = "Tumasyan, Armen and others",
	collaboration = "CMS",
	title = "{A portrait of the Higgs boson by the CMS experiment ten years after the discovery}",
	eprint = "2207.00043",
	archivePrefix = "arXiv",
	primaryClass = "hep-ex",
	doi = "10.1038/s41586-022-04892-x",
	journal = "Nature",
	volume = "607",
	number = "7917",
	pages = "60--68",
	year = "2022"
}

@article{Cepeda:2019klc,
	author = "Cepeda, M. and others",
	title = "{Report from Working Group 2: Higgs Physics at the HL-LHC and HE-LHC}",
	eprint = "1902.00134",
	archivePrefix = "arXiv",
	primaryClass = "hep-ph",
	reportNumber = "CERN-LPCC-2018-04",
	doi = "10.23731/CYRM-2019-007.221",
	journal = "CERN Yellow Rep. Monogr.",
	volume = "7",
	pages = "221--584",
	year = "2019"
}

@article{deBlas:2019rxi,
	author = "de Blas, J. and others",
	title = "{Higgs Boson Studies at Future Particle Colliders}",
	eprint = "1905.03764",
	archivePrefix = "arXiv",
	primaryClass = "hep-ph",
	reportNumber = "DESY-19-079",
	doi = "10.1007/JHEP01(2020)139",
	journal = "JHEP",
	volume = "01",
	pages = "139",
	year = "2020"
}

@article{dEnterria:2021xij,
	author = "d'Enterria, David and Poldaru, Andres and Wojcik, George",
	title = "{Measuring the electron Yukawa coupling via resonant s-channel Higgs production at FCC-ee}",
	eprint = "2107.02686",
	archivePrefix = "arXiv",
	primaryClass = "hep-ex",
	doi = "10.1140/epjp/s13360-021-02204-2",
	journal = "Eur. Phys. J. Plus",
	volume = "137",
	number = "2",
	pages = "201",
	year = "2022"
}

@article{FCC:2025lpp,
	author = "Benedikt, M. and others",
	collaboration = "FCC",
	title = "{Future Circular Collider Feasibility Study Report: Volume 1, Physics, Experiments, Detectors}",
	eprint = "2505.00272",
	archivePrefix = "arXiv",
	primaryClass = "hep-ex",
	reportNumber = "CERN-FCC-PHYS-2025-0002",
	doi = "10.1140/epjc/s10052-025-15077-x",
	journal = "Eur. Phys. J. C",
	volume = "85",
	number = "12",
	pages = "1468",
	year = "2025",
	note = "[Erratum: Eur.Phys.J.C 86, 844 (2026)]"
}

@article{Crivellin:2021rbq,
	author = "Crivellin, Andreas and Hoferichter, Martin",
	title = "{Consequences of chirally enhanced explanations of (g {\ensuremath{-}} 2)$_{\mu}$ for h {\textrightarrow} $\mu\mu$ and Z {\textrightarrow} $\mu\mu$}",
	eprint = "2104.03202",
	archivePrefix = "arXiv",
	primaryClass = "hep-ph",
	doi = "10.1007/JHEP07(2021)135",
	journal = "JHEP",
	volume = "07",
	pages = "135",
	year = "2021",
	note = "[Erratum: JHEP 10, 030 (2022)]"
}

@article{Freitas:2014pua,
	author = "Freitas, Ayres and Lykken, Joseph and Kell, Stefan and Westhoff, Susanne",
	title = "{Testing the Muon g-2 Anomaly at the LHC}",
	eprint = "1402.7065",
	archivePrefix = "arXiv",
	primaryClass = "hep-ph",
	doi = "10.1007/JHEP05(2014)145",
	journal = "JHEP",
	volume = "05",
	pages = "145",
	year = "2014",
	note = "[Erratum: JHEP 09, 155 (2014)]"
}

@article{Athron:2021iuf,
	author = {Athron, Peter and Bal{\'a}zs, Csaba and Jacob, Douglas H. J. and Kotlarski, Wojciech and St{\"o}ckinger, Dominik and St{\"o}ckinger-Kim, Hyejung},
	title = "{New physics explanations of $a_\mu$ in light of the FNAL muon $g-2$ measurement}",
	eprint = "2104.03691",
	archivePrefix = "arXiv",
	primaryClass = "hep-ph",
	doi = "10.1007/JHEP09(2021)080",
	journal = "JHEP",
	volume = "09",
	pages = "080",
	year = "2021"
}

@article{Athron:2025ets,
	author = {Athron, Peter and M{\"o}hling, Kilian and St{\"o}ckinger, Dominik and St{\"o}ckinger-Kim, Hyejung},
	title = "{The muon magnetic moment and physics beyond the standard model}",
	eprint = "2507.09289",
	archivePrefix = "arXiv",
	primaryClass = "hep-ph",
	reportNumber = "CPTNP-2025-023",
	doi = "10.1016/j.ppnp.2025.104225",
	journal = "Prog. Part. Nucl. Phys.",
	volume = "148",
	pages = "104225",
	year = "2026"
}

@article{Fan:2022eto,
	author = "Fan, X. and Myers, T. G. and Sukra, B. A. D. and Gabrielse, G.",
	title = "{Measurement of the Electron Magnetic Moment}",
	eprint = "2209.13084",
	archivePrefix = "arXiv",
	primaryClass = "physics.atom-ph",
	doi = "10.1103/PhysRevLett.130.071801",
	journal = "Phys. Rev. Lett.",
	volume = "130",
	number = "7",
	pages = "071801",
	year = "2023"
}

@article{ATLAS:2019old,
	author = "Aad, Georges and others",
	collaboration = "ATLAS",
	title = "{Search for the Higgs boson decays $H \to ee$ and $H \to e\mu$ in $pp$ collisions at $\sqrt{s} = 13$ TeV with the ATLAS detector}",
	eprint = "1909.10235",
	archivePrefix = "arXiv",
	primaryClass = "hep-ex",
	doi = "10.1016/j.physletb.2019.135148",
	journal = "Phys. Lett. B",
	volume = "801",
	pages = "135148",
	year = "2020"
}

@article{CMS:2022urr,
	author = "Tumasyan, Armen and others",
	collaboration = "CMS",
	title = "{Search for the Higgs boson decay to a pair of electrons in proton-proton collisions at $\sqrt{s}$ = 13 TeV}",
	eprint = "2208.00265",
	archivePrefix = "arXiv",
	primaryClass = "hep-ex",
	doi = "10.1016/j.physletb.2023.137783",
	journal = "Phys. Lett. B",
	volume = "846",
	pages = "137783",
	year = "2023"
}

@article{ATLAS:2022ryk,
	author = "Aad, Georges and others",
	collaboration = "ATLAS",
	title = "{Observation of the $\gamma\gamma\to\tau\tau$ Process in Pb+Pb Collisions and Constraints on the $\tau$-Lepton Anomalous Magnetic Moment with the ATLAS Detector}",
	eprint = "2204.13478",
	archivePrefix = "arXiv",
	primaryClass = "hep-ex",
	doi = "10.1103/PhysRevLett.131.151802",
	journal = "Phys. Rev. Lett.",
	volume = "131",
	number = "15",
	pages = "151802",
	year = "2023"
}

@article{CMS:2024qjo,
	author = "Hayrapetyan, Aram and others",
	collaboration = "CMS",
	title = "{Observation of $\gamma\gamma\to\tau\tau$ in proton-proton collisions and limits on the anomalous electromagnetic moments of the $\tau$ lepton}",
	eprint = "2406.03975",
	archivePrefix = "arXiv",
	primaryClass = "hep-ex",
	doi = "10.1088/1361-6633/ad6fcb",
	journal = "Rept. Prog. Phys.",
	volume = "87",
	number = "10",
	pages = "107801",
	year = "2024"
}

@article{Belle:2021ybo,
	author = "Inami, K. and others",
	collaboration = "Belle",
	title = "{An improved search for the electric dipole moment of the $\tau$ lepton}",
	eprint = "2108.11543",
	archivePrefix = "arXiv",
	primaryClass = "hep-ex",
	reportNumber = "Belle Preprint 2021-22, KEK Preprint 2021-26",
	doi = "10.1007/JHEP04(2022)110",
	journal = "JHEP",
	volume = "04",
	pages = "110",
	year = "2022"
}

@article{ParticleDataGroup:2026mpi,
	author = "Takahashi, F. and others",
	collaboration = "Particle Data Group",
	title = "{Review of Particle Physics}",
	doi = "10.1142/s0217751x26300115",
	journal = "Int. J. Mod. Phys. A",
	volume = "41",
	number = "22",
	pages = "2630011",
	year = "2026"
}

@article{Parker:2018vye,
	author = {Parker, Richard H. and Yu, Chenghui and Zhong, Weicheng and Estey, Brian and M{\"u}ller, Holger},
	title = "{Measurement of the fine-structure constant as a test of the Standard Model}",
	eprint = "1812.04130",
	archivePrefix = "arXiv",
	primaryClass = "physics.atom-ph",
	doi = "10.1126/science.aap7706",
	journal = "Science",
	volume = "360",
	pages = "191",
	year = "2018"
}

@article{Morel:2020dww,
	author = {Morel, L{\'e}o and Yao, Zhibin and Clad{\'e}, Pierre and Guellati-Kh{\'e}lifa, Sa{\"\i}da},
	title = "{Determination of the fine-structure constant with an accuracy of 81 parts per trillion}",
	doi = "10.1038/s41586-020-2964-7",
	journal = "Nature",
	volume = "588",
	number = "7836",
	pages = "61--65",
	year = "2020"
}

@article{Muong-2:2008ebm,
	author = "Bennett, G. W. and others",
	collaboration = "Muon (g-2)",
	title = "{An Improved Limit on the Muon Electric Dipole Moment}",
	eprint = "0811.1207",
	archivePrefix = "arXiv",
	primaryClass = "hep-ex",
	doi = "10.1103/PhysRevD.80.052008",
	journal = "Phys. Rev. D",
	volume = "80",
	pages = "052008",
	year = "2009"
}

@article{ParticleDataGroup:2024cfk,
	author = "Navas, S. and others",
	collaboration = "Particle Data Group",
	title = "{Review of particle physics}",
	doi = "10.1103/PhysRevD.110.030001",
	journal = "Phys. Rev. D",
	volume = "110",
	number = "3",
	pages = "030001",
	year = "2024"
}

@article{Sakurai:2022tbk,
	author = "Sakurai, Mikio and others",
	title = "{muEDM: Towards a Search for the Muon Electric Dipole Moment at PSI Using the Frozen-spin Technique}",
	eprint = "2201.06561",
	archivePrefix = "arXiv",
	primaryClass = "hep-ex",
	doi = "10.7566/JPSCP.37.020604",
	journal = "JPS Conf. Proc.",
	volume = "37",
	pages = "020604",
	year = "2022"
}

@article{Crivellin:2021spu,
	author = "Crivellin, Andreas and Hoferichter, Martin and Roney, J. Michael",
	title = "{Toward testing the magnetic moment of the tau at one part per million}",
	eprint = "2111.10378",
	archivePrefix = "arXiv",
	primaryClass = "hep-ph",
	reportNumber = "PSI-PR-21-27, ZU-TH 56/21",
	doi = "10.1103/PhysRevD.106.093007",
	journal = "Phys. Rev. D",
	volume = "106",
	number = "9",
	pages = "093007",
	year = "2022"
}

@inproceedings{USBelleIIGroup:2022qro,
	author = "Asner, D. M. and others",
	collaboration = "US Belle II Group, Belle II/SuperKEKB e- Polarization Upgrade Working Group",
	title = "{Snowmass 2021 White Paper on Upgrading SuperKEKB with a Polarized Electron Beam: Discovery Potential and Proposed Implementation}",
	booktitle = "{Snowmass 2021}",
	eprint = "2205.12847",
	archivePrefix = "arXiv",
	primaryClass = "physics.acc-ph",
	month = "5",
	year = "2022"
}

@article{Collins:2016aya,
	author = "Collins, John C. and Vermaseren, J. A. M.",
	title = "{Axodraw Version 2}",
	eprint = "1606.01177",
	archivePrefix = "arXiv",
	primaryClass = "cs.OH",
	month = "5",
	year = "2016"
}
	
\end{document}